\documentclass[12pt,a4paper]{article}
\usepackage{graphicx}
\usepackage{amssymb,amsfonts,amsmath}
\usepackage{color}

\usepackage{cite}

\makeatletter
\renewcommand{\@biblabel}[1]{}
\makeatother

\begin{document}

\newcommand{\EQ}{Eq.~}
\newcommand{\EQS}{Eqs.~}
\newcommand{\FIG}{Fig.~}
\newcommand{\FIGS}{Figs.~}
\newcommand{\TAB}{Tab.~}
\newcommand{\TABS}{Tabs.~}
\newcommand{\SEC}{Sec.~}
\newcommand{\SECS}{Secs.~}

\setlength{\baselineskip}{0.77cm}

\title{Upstream reciprocity in heterogeneous networks}
\bigskip

\author{Akio Iwagami${}^{1}$ and Naoki Masuda${}^{2,3*}$\\
\ \\
\ \\
${}^1$ Faculty of Engineering,
The University of Tokyo,\\
7-3-1, Hongo, Bunkyo, Tokyo 113-8656, Japan\\
\ \\
${}^{2}$ 
Graduate School of Information Science and Technology,\\
The University of Tokyo,\\
7-3-1 Hongo, Bunkyo, Tokyo 113-8656, Japan
\ \\
${}^3$ PRESTO, Japan Science and Technology Agency,\\
4-1-8 Honcho, Kawaguchi, Saitama 332-0012, Japan
\ \\
$^*$ Author for correspondence (masuda@mist.i.u-tokyo.ac.jp)}

\maketitle

\newpage

\begin{abstract}
\setlength{\baselineskip}{0.77cm} 

\end{abstract}
Many mechanisms for the emergence and maintenance of altruistic behavior in social dilemma situations have been proposed. Indirect reciprocity is one such mechanism, where other-regarding actions of a player are eventually rewarded by other players with whom the original player has not interacted. The upstream reciprocity (also called generalized indirect reciprocity) is a type of indirect reciprocity and represents the concept that those helped by somebody will help other unspecified players. In spite of the evidence for the enhancement of helping behavior by upstream reciprocity in rats and humans, theoretical support for this mechanism is not strong. In the present study, we numerically investigate upstream reciprocity in heterogeneous contact networks, in which the players generally have different number of neighbors.  We show that heterogeneous networks considerably enhance cooperation in a game of upstream reciprocity. In heterogeneous networks, the most generous strategy, by which a player helps a neighbor on being helped and in addition initiates helping behavior, first occupies hubs in a network and then disseminates to other players. The scenario to achieve enhanced altruism resembles that seen in the case of the Prisoner's Dilemma game in heterogeneous networks.
\newpage

\section{Introduction}\label{sec:introduction}

The mechanism for evolution and maintenance of altruism when
egoistic behavior is apparently more advantageous
has been a target of intensive studies.  Among the
many viable mechanisms proposed, we focus on
indirect reciprocity, which refers to
the concept that a cooperative player
is helped by others with whom she/he
has not interacted. Cooperative behavior is indirectly 
rewarded by way of chains of helping behavior of various players.
There are two types of indirect reciprocity:
downstream reciprocity and upstream reciprocity \cite{Nowak05nat}.  In
downstream reciprocity, a player witnesses the
behavior of other players as a
third party. The observing player will assign a good reputation to
player $X$ if player $X$ helps others. When a situation arises where
this observer interacts with
player $X$ in the future, the observer will probably
help $X$ if and only if $X$ has a good reputation.
A player must establish a good
reputation by helping others prior to being helped by other
anonymous players. 
The downstream reciprocity is observed in behavioral experiments
\cite{Wedekind00,Milinski02} and is firmly based on
the theory of evolutionary games
\cite{Nowak98nat,Nowak98jtb,Leimar01,Brandt04,OhtsukiIwasa04jtb,OhtsukiIwasa06jtb_eight}.

In upstream reciprocity, the players first get help from other
players.  If the recipient complies with upstream reciprocity, then
she/he helps another unspecified player.  Theoretically, evolution of
cooperation based on upstream reciprocity is considered to be
difficult. In numerical simulations, cooperation is achieved only when
the size of the interaction group is small
\cite{Boyd89socn,Pfeiffer05}. An analytical study showed that upstream
reciprocity enables evolution of cooperation only in combination with
another mechanism such as direct reciprocity (\textit{i.e.}, repeated
interaction between the same players) or spatial reciprocity
(\textit{i.e.}, interaction between players on a one-dimensional
lattice) \cite{Nowak07}. 
However, upstream reciprocity has been
observed  in behavioral experiments conducted on humans.
A player that has received a help from another player
has increased the propensity to help
an anonymous partner in variants of 
the trust game \cite{Dufwenberg01,Greiner05,Stanca09}.
Those who are helped by somebody in advance
tend to help another partner filling in a
tedious survey in
laboratory behavioral experiments
\cite{Bartlett06}. 
Upstream reciprocity has also been observed in rats. Rats
trained to pull a stick to
deliver food tend to pull the stick to help another rat
after receiving food via a help from a conspecific
\cite{Rutte07}.
Therefore, theoretically assessing the conditions
under which upstream reciprocity is feasible will help us gain a
better understanding of the evolution of cooperation in social dilemma
situations.

In this study, we examine the effect of a property of contact networks
on upstream reciprocity.
A fundamental characteristic of many social networks is that
the number of contacts of a node, which we call the degree, 
has a right-skewed distribution. In particular,
scale-free networks, \textit{i.e.}, networks with
power-law degree distributions are widely found
(e.g., Newman, 2003).
In social networks relevant to evolutionary games, scale-free networks
have been found in, for example, email social networks
\cite{Ebel02pre_email,Newman02pre_email}. 
Although other social networks
do not exhibit degree distributions that are as right skewed as
the power-law distribution,
their degree distributions are considerably heterogeneous
\cite{Eubank04,Lusseau04,Kossinets06,Onnela07pnas}.
We investigate the effect of heterogeneous degree distributions
on the possible evolution of cooperation based on upstream reciprocity.

We show that upstream reciprocity enhances altruistic behavior of
players that are placed in heterogeneous contact networks such as
scale-free networks.
The mechanism found in our study has resemblance to that
for enhanced cooperation shown in the Prisoner's Dilemma in heterogeneous 
networks \cite{Duran05,Santos05prl,Santos06pnas,Santos06jeb}, which
we will discuss in \SEC\ref{sec:discussion}.

\section{Model}

\subsection{Networks}

\begin{figure}
\begin{center}
\includegraphics[clip, height=6cm,width=6cm]{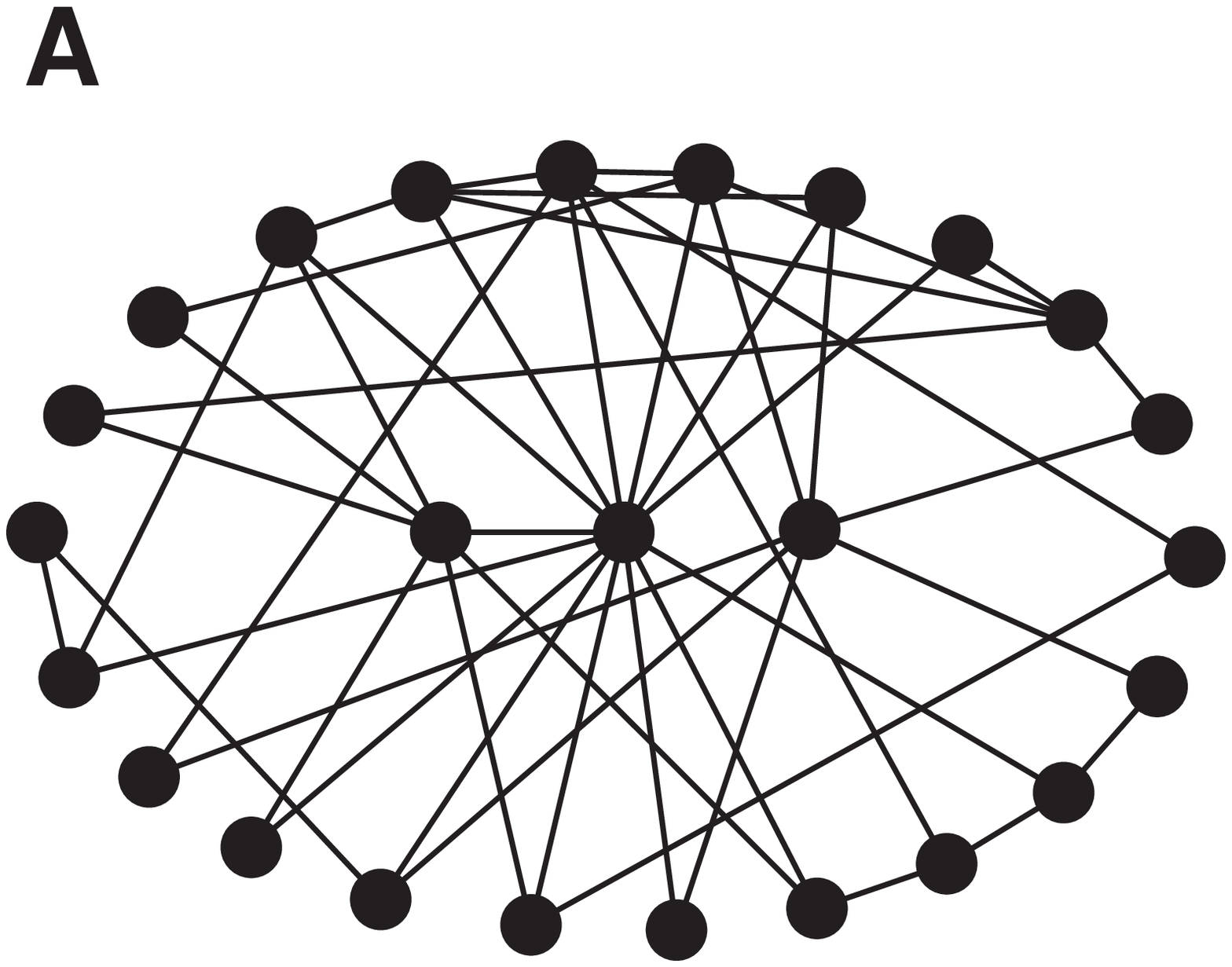}
\includegraphics[clip, height=6cm,width=6cm]{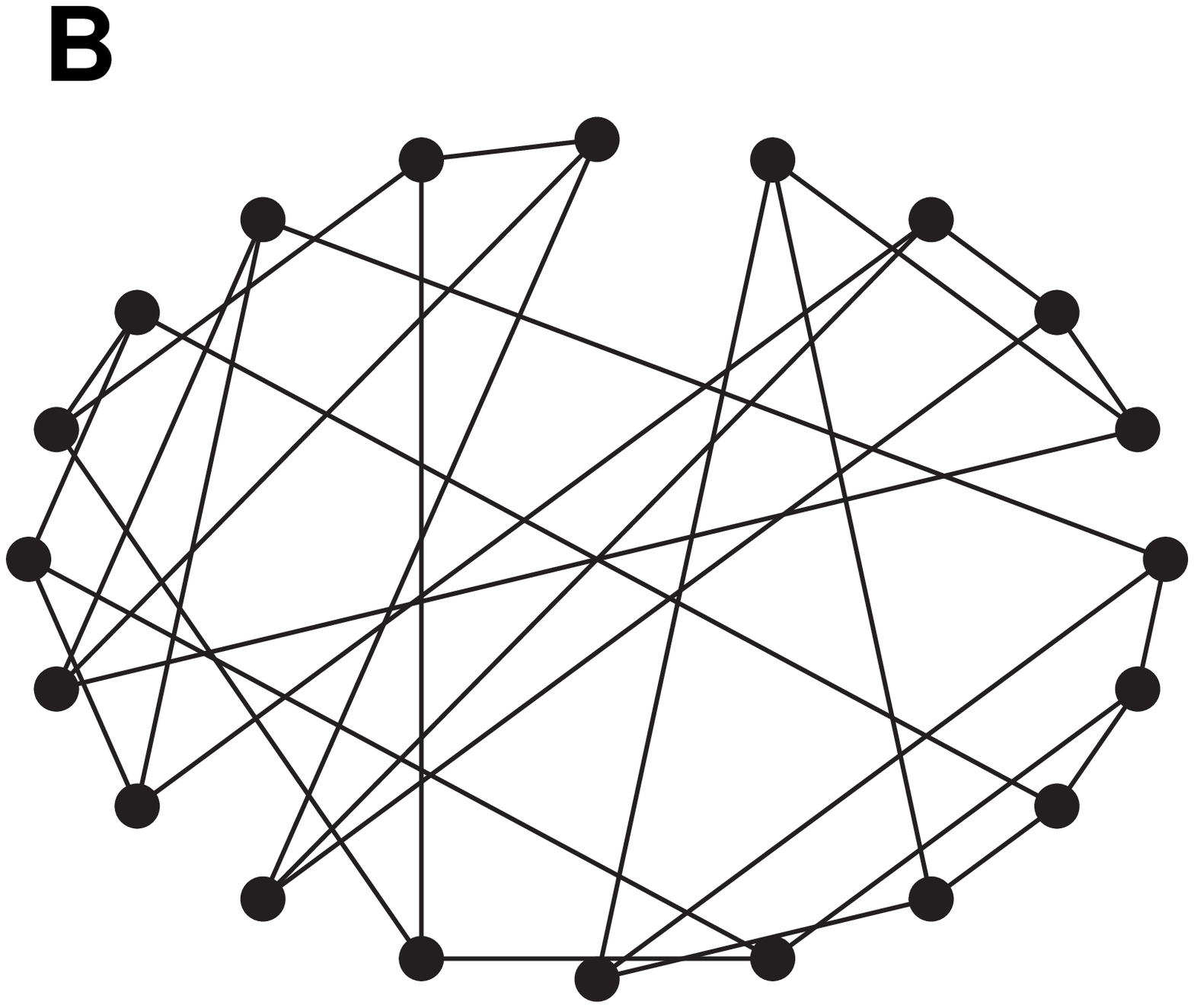}
\includegraphics[clip, height=4.5cm,width=6cm]{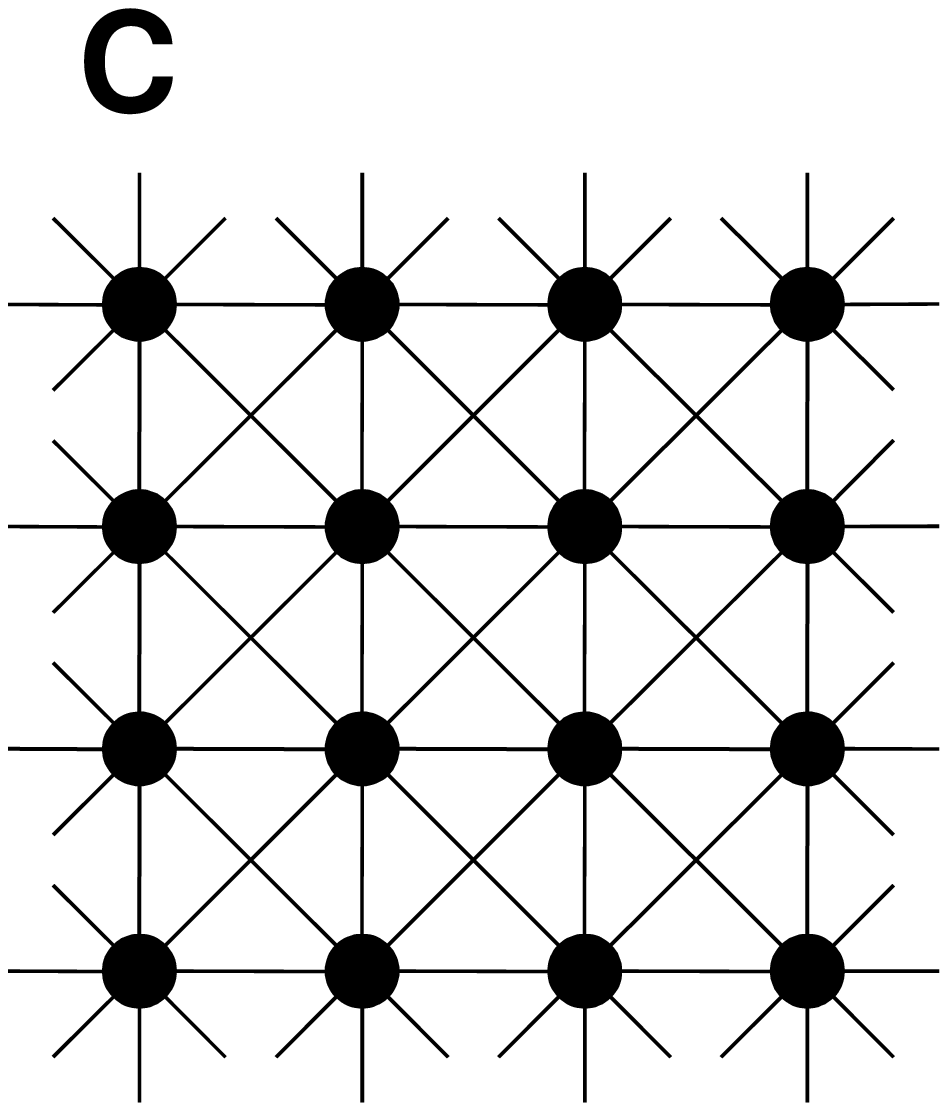}
\includegraphics[clip, height=4.5cm,width=6cm]{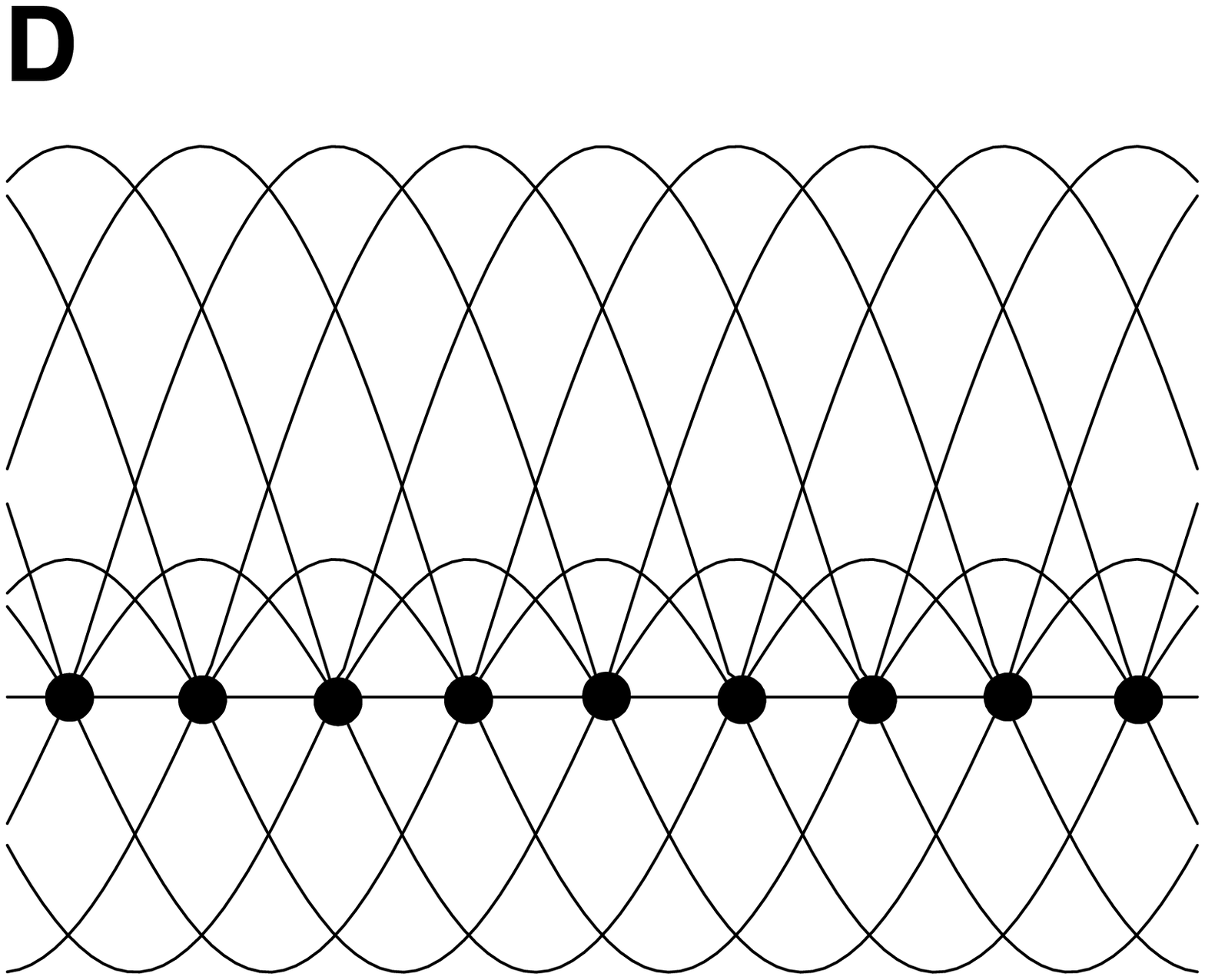}
\includegraphics[clip, height=1.5cm,width=6cm]{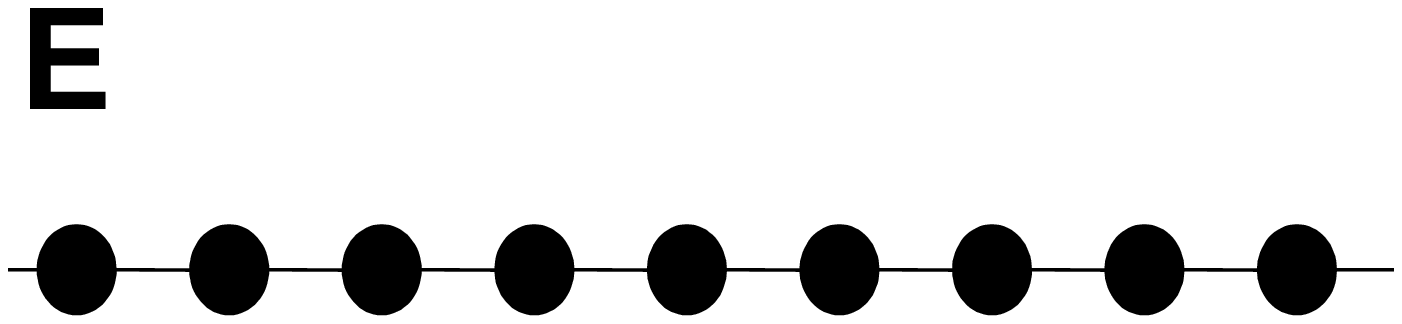}
\caption{Architecture of networks.
(A) Scale-free network, (B) regular random graph,
(C) square lattice, (D) extended cycle, and (E) cycle.}
\label{fig:network}
\end{center}
\end{figure}

Consider a contact network with a
population of $N=10000$ players.  As a
model of heterogeneous network, we use the scale-free network
generated by the Barab\'{a}si--Albert algorithm \cite{Barabasi99sci}
(\FIG\ref{fig:network}A).  To generate the scale-free network, we start
with the complete graph of $2m+1$ nodes (\textit{i.e.}, each pair of
nodes is connected by an edge).  Then,
we add nodes with degree $m$ one-by-one according to the
so-called linear preferential attachment; the probability that an
already existing node $v_i$ forms an edge with a newly introduced
node is proportional to the degree $k_i$. Multiple edges
(\textit{i.e.}, more than one edge connecting a pair of nodes) are
disallowed. In the generated network, the degree follows the power-law
distribution $p(k)\propto k^{-3}$ with a lower cutoff at $k=m$ and the
mean degree of $\left<k\right>=2m$ \cite{Barabasi99sci}.  We use
$\left<k\right>=8$, \textit{i.e.}, $m=4$, unless otherwise stated.

For comparison, we also use four other types of networks.  One is the
regular random graph, which is constructed from the configuration
model \cite{Newman03siam} (\FIG\ref{fig:network}B).  To generate a
network, we attach $\left<k\right>$ stubs, or half edges, to each
node.  Then, we randomly select two nodes with the equal selection
probability and connect them. These two nodes consume one stub
each. We repeat this procedure until all stubs are exhausted at all
nodes.  If the generated network is disconnected or contains 
self-loops or multiple edges, we discard the network and start the entire
procedure all over again. Although its mean degree is small, the
regular random graph represents a well-mixed population in which
cooperation is not easily enhanced by upstream reciprocity
\cite{Boyd89socn,Nowak07}.

In the square lattice,
$N=10000$ nodes are placed on the square with a
linear length of $\sqrt{N}=100$.
Each node is connected to eight nodes situated in a so-called
Moore neighborhood (\FIG\ref{fig:network}C).
We adopt the periodic boundary condition.

The extended cycle is a one-dimensional network, where the nodes are placed
on a ring. Each node is connected to $\left<k\right>/2$ nearest nodes on each
side, as shown in \FIG\ref{fig:network}D.

The scale-free network, the regular random graph, the square lattice,
and the extended cycle have $\left<k\right>=8$ unless otherwise
stated.  Therefore, we can compare the effects of different types of
networks without having to account for the possible influence of
$\left<k\right>$.  We also set $\left<k\right>=6$ and
$\left<k\right>=14$ in some of the following numerical
simulations to confirm
the robustness of the results with respect to
$\left<k\right>$.

The final type of network used is the cycle in which each node on a
ring is connected to a single nearest node on each side such that
$\left<k\right>=2$ (\FIG\ref{fig:network}E).  We use the cycle to
compare our numerical results with the previously reported theoretical
results \cite{Nowak07}. In contrast to the well-mixed population, the
infinite one-dimensional chain network with $\left<k\right>=2$ enables
upstream reciprocity because it exhibits spatial reciprocity.  Spatial
reciprocity is a general mechanism for evolution of cooperation in
social dilemma games; cooperative players are clustered in a network to
help each other and resist the invasion by egoistic players
\cite{Axelrod84,Nowak92}. Such clustering is possible when the size of
the boundary of a cluster is small relative to the number of players
in the cluster. This situation is expected the most in the cycle and
to a certain extent in the extended cycle and the square lattice;
however, it is not expected in the 
Barab\'{a}si--Albert scale-free network and the regular
random graph.

\subsection{Game of upstream reciprocity: rule and payoff}

A single game of upstream reciprocity \cite{Nowak07}, which is
motivated by experimental evidence and previous theoretical work explained in \SEC\ref{sec:introduction}, is described as follows.
First, a player $v_i$ ($1\le i\le N$) is selected.
Player $v_i$ may initiate a chain of helping behavior. If $v_i$ does
so, $v_i$ bears the cost $c$ and selects one of its neighbors at an
equal selection probability of $1/k_i$, where $k_i$ is the degree of
$v_i$. The selected neighbor, denoted by $v_j$, receives the payoff
$b$. We assume $b>c>0$ so that the game represents a social dilemma; a
single act of help increases the average payoff of the entire
population by $(b-c)/N$, while each player is better off by not
helping other players. Without loss of generality, we set $c=1$.

$v_j$ may not continue the chain of helping behavior. In such a case, the chain of
cooperation terminates, and the payoffs for $v_i$, $v_j$, and
$v_{i^{\prime}}$ ($i^{\prime}\neq i,j$) are equal to $-c$, $b$, and 0,
respectively. However, if $v_j$ does pass on the helping action,
$v_j$ selects one
of its neighbors at a probability of
$1/k_j$ and bears the cost $c$. The
selected neighbor receives $b$. The chain of helping behavior
continues until a recipient of help terminates the chain. Note that
a chain of cooperation may traverse the same players more than once.

\subsection{Strategies}

On the basis of a previous study \cite{Nowak07}, we specify the
strategy of each player $v_i$ ($1\le i\le N$)
using two parameters. The first parameter $p_i$ ($0\le p_i\le 1$)
denotes the probability that $v_i$ passes on the helping action
to a randomly selected neighbor after receiving it
from a neighbor. The second parameter $q_i$ ($0\le q_i\le 1$)
denotes the probability that $v_i$ initiates the helping action.
A larger $p_i$ or $q_i$ implies that player $v_i$
is more cooperative.

We consider the following four strategies that were introduced by
Nowak and Roch (2007): 
\begin{itemize}
\item
Classical defector (CD) is defined by
$p_i=0$ and $q_i=0$. CD neither initiates nor passes on the help.
It is the most egoistic strategy.

\item
Classical cooperator (CC) is defined by $p_i=0$ and $q_i=1$.  CC
spontaneously initiates the chain of helping behavior
but does not react to the cooperation
that it receives from a neighbor.  CC does not contribute to upstream
reciprocity, even though CC is cooperative to some extent.

\item
Generous cooperator (GC) is defined by $p_i=0.8$ and $q_i=1$.  GC
initiates the helping behavior
and passes on the helping action with a high probability.
It is the most cooperative strategy. We are concerned with the
possibility that heterogeneous networks enhance the fraction of GCs in
a population.

\item
Passer-on (PO) is defined by $p_i=0.8$ and $q_i=0$. PO does not initiate
the helping behavior
but passes on the helping action with a high probability. Although
PO is less cooperative than GC, it contributes to the
upstream reciprocity.

\end{itemize}

In the case of GC and PO, we set $p_i=0.8$ instead of $p_i=1$.
This is to prevent
a chain of helping behavior from continuing indefinitely if
the population consists of only GC and PO.
This choice of $p_i$ is arbitrary. To verify the robustness of 
our results with respect to the value of $p_i$, we will carry out
some of the following numerical simulations with $p_i=0.7$ and $p_i=0.9$.

\subsection{Update rule}\label{sub:update rule}

We principally use the deterministic update rule, which is described in the
following. The numerical results do not qualitatively change on using
relatively realistic stochastic rules, as shown in
\SECS\ref{sub:GC CD} and \ref{sub:4 strategies}.

We refer to time in the evolutionary dynamics
as a round and denote it
by $t$ ($=0, 1, 2,\ldots$).
One round consists of $N$ chains of
helping behavior, and one chain is initiated by
each player. Note that a chain is considered to be empty
if the initial player does not help a neighbor, which occurs
for CD and PO.  The
one-round payoff of player $v_i$ is defined as the sum of the
payoffs gained by $v_i$ in $N$ chains of cooperation.  The payoff
that $v_i$ gains in a round is equal to $b\, \times$ (the frequency
at which the chains are brought to $v_i$) $-$ $c\, \times$ (the frequency
at which the chains are passed from $v_i$ without being terminated).

At the end of each round, the strategies of $N_{\rm u}$
out of the $N=10000$
players are updated synchronously. 
Unless otherwise stated, we set $N_{\rm u}=200$.
We also set $N_{\rm u}=20$ and $N_{\rm u}=2000$ in some of the
following numerical simulations to examine the robustness of the results
with respect to $N_{\rm u}$.
We randomly and independently
select $N_{\rm u}$ players from the population with equal
probability. In the deterministic update rule that we mostly
use in this paper, for each
selected player $v_i$, the neighbor with the largest
payoff, which is denoted by $v_j$, is selected. 
If the payoff of $v_j$ is larger than that
of $v_i$, $v_i$ will copy the strategy of $v_j$. If there are
more than one neighbors with the same largest payoff, we select one
of them randomly with equal probability. After tentatively
determining $N_{\rm u}$ copying events, we
replace the strategies of the selected nodes simultaneously.  We do not
assume mutation. This marks the end of one round.

One run lasts until a quasistationary state is attained
or the unanimity of one
strategy is almost achieved.  Specifically, we
set the number of rounds to 20000 in the case of
the scale-free network, the
regular random graph, and the square lattice. 
In the case of the extended cycle and the cycle, the number of rounds
is equal to 140000.

\section{Results}\label{sec:results}

\subsection{GC versus CD}\label{sub:GC CD}

When a player passes on the received help to a neighbor, a neighbor is
randomly selected as recipient with equal probability. A chain of
helping behavior is equivalent to a simple random walk with random
termination.  If $p_i=1$ ($1\le i\le N$), the random walk may continue
forever. In this hypothetical situation, the payoff that player $i$
receives is proportional to the stationary density of the random walk.
In any undirected network, the stationary density of the simple random
walk is proportional to the degree (\textit{e.g.},
Noh and Rieger, 2004).
This relation
roughly holds true for uncorrelated networks even in the presence of some
absorbing nodes at which the random walk terminates \cite{Noh04prl}.
Therefore, we expect that the number of times that the chain of helping
behavior reaches a given 
node is roughly proportional to the degree. Because
a single passage of chain contributes to the payoff $b-c>0$, the
payoff per round for each player is roughly proportional to the
degree.

\begin{figure}
\begin{center}
\includegraphics[height=6cm,width=8cm]{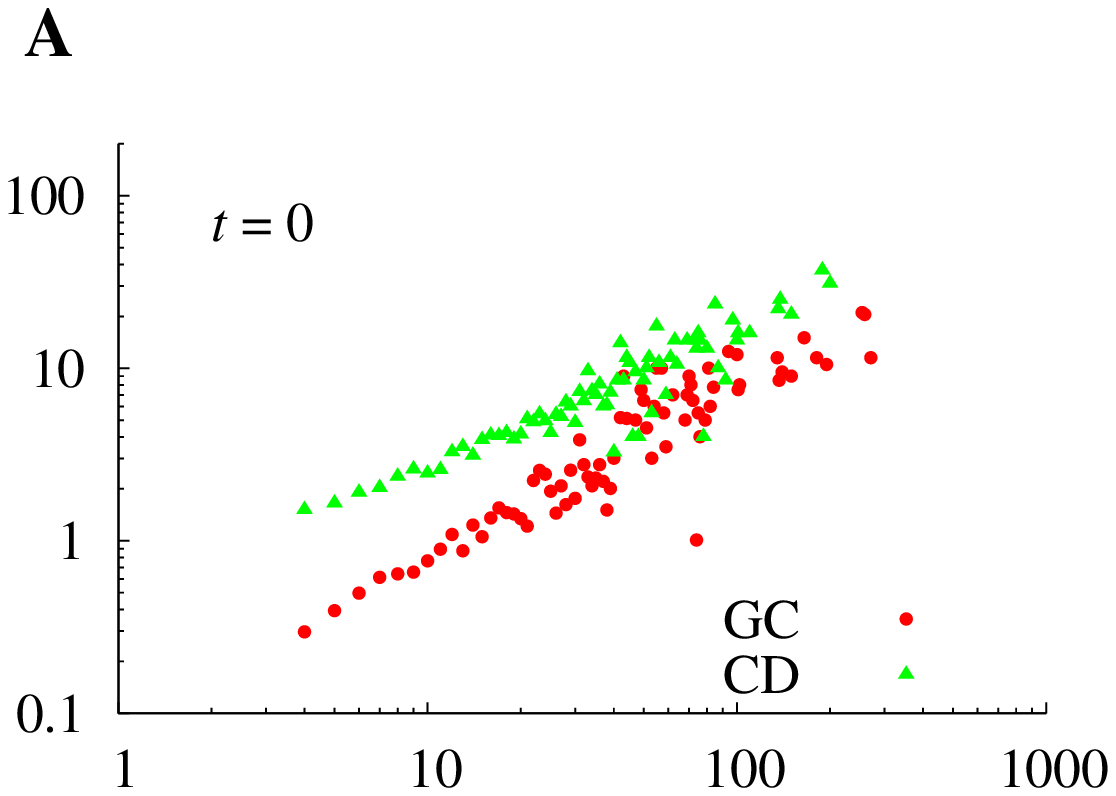}
\includegraphics[height=6cm,width=8cm]{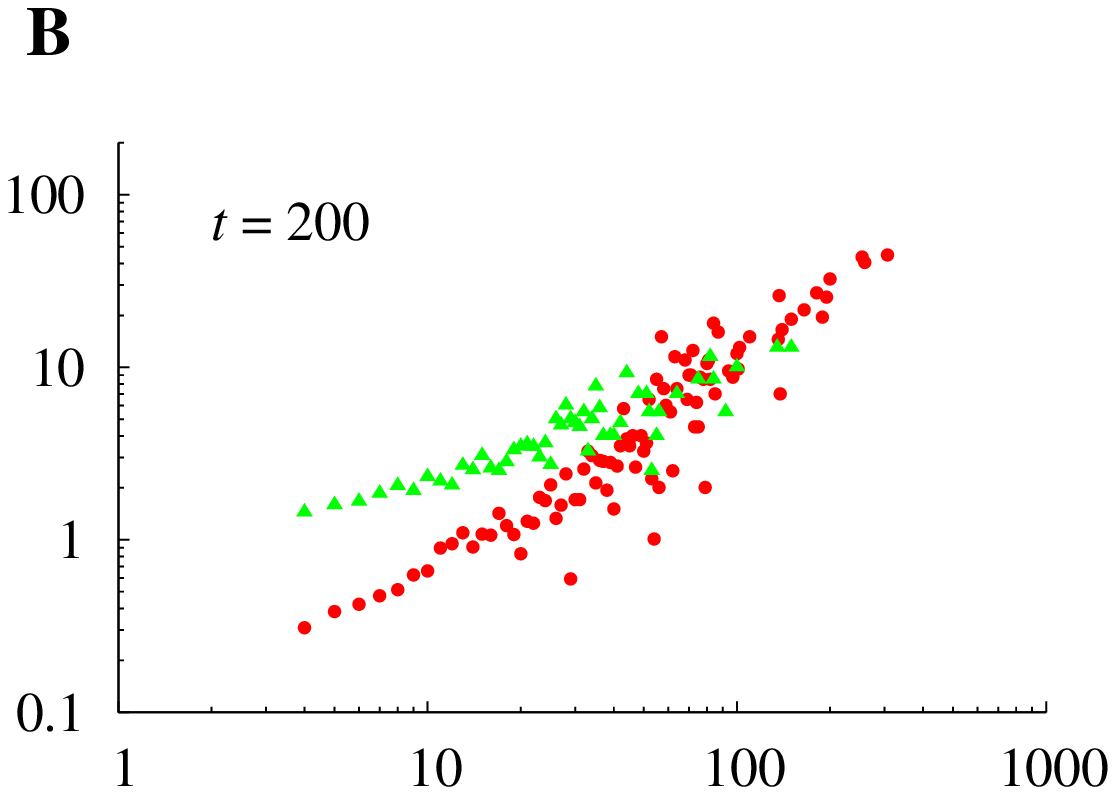}
\includegraphics[height=6cm,width=8cm]{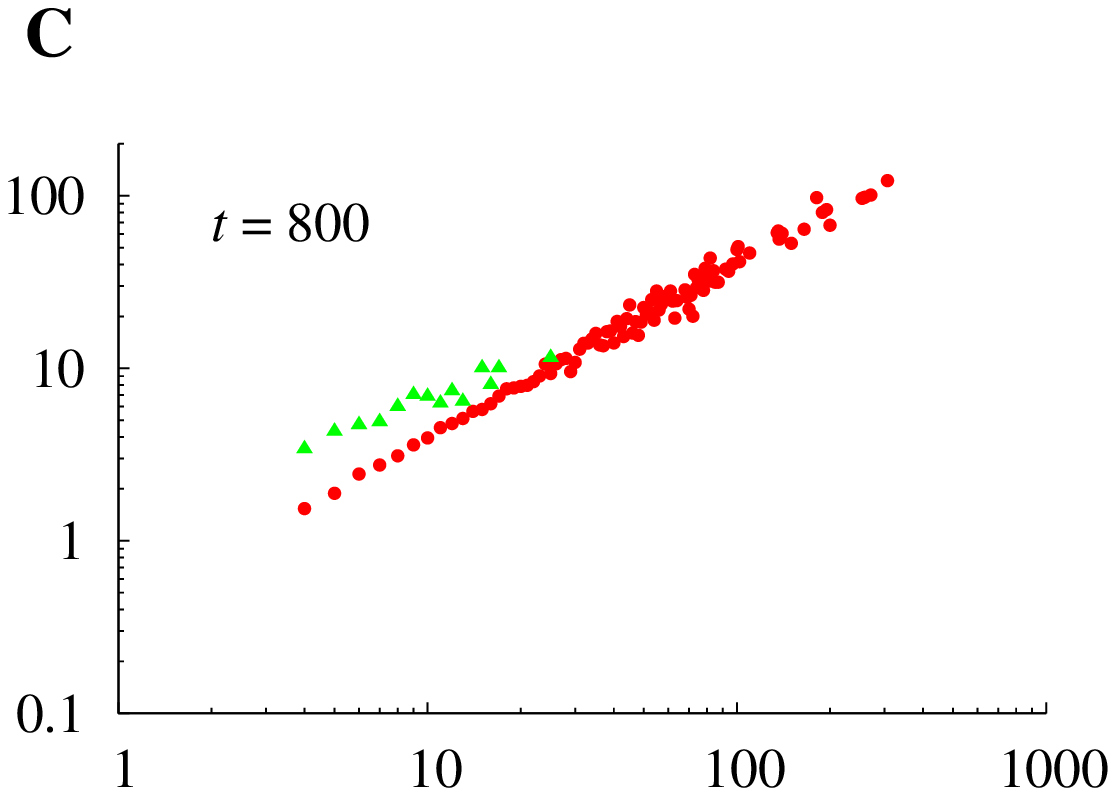}
\includegraphics[height=6cm,width=8cm]{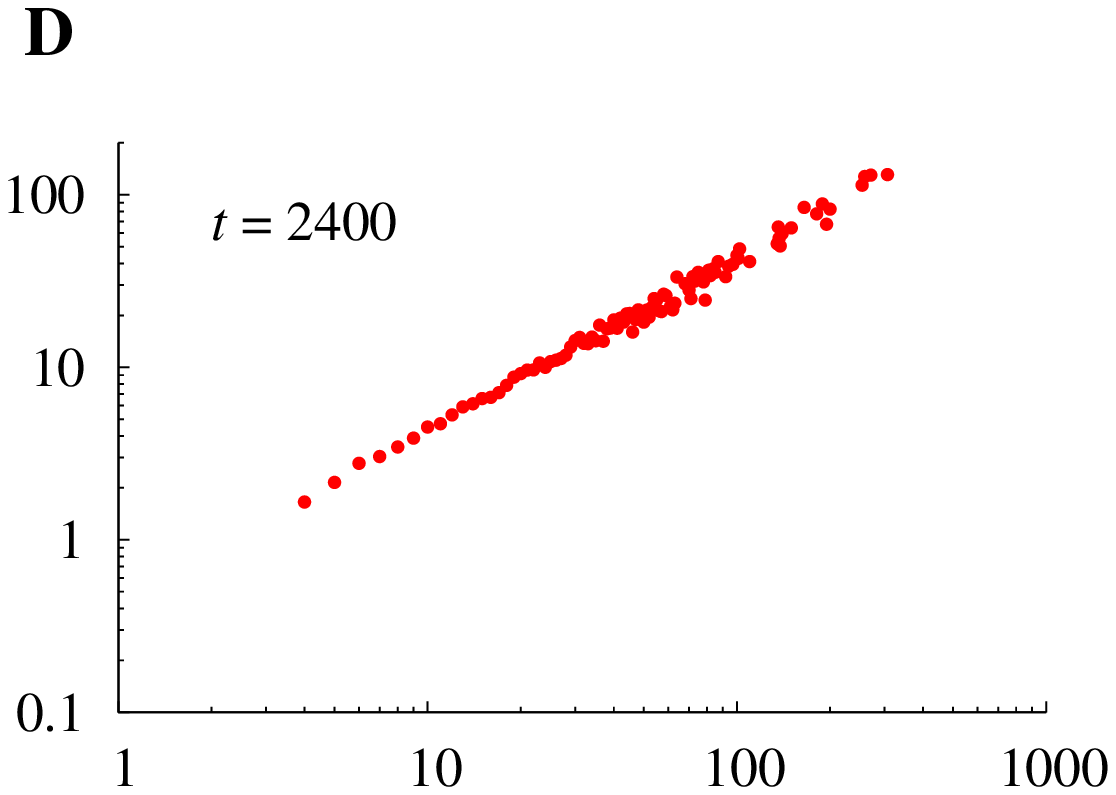}
\includegraphics[height=6cm,width=8cm]{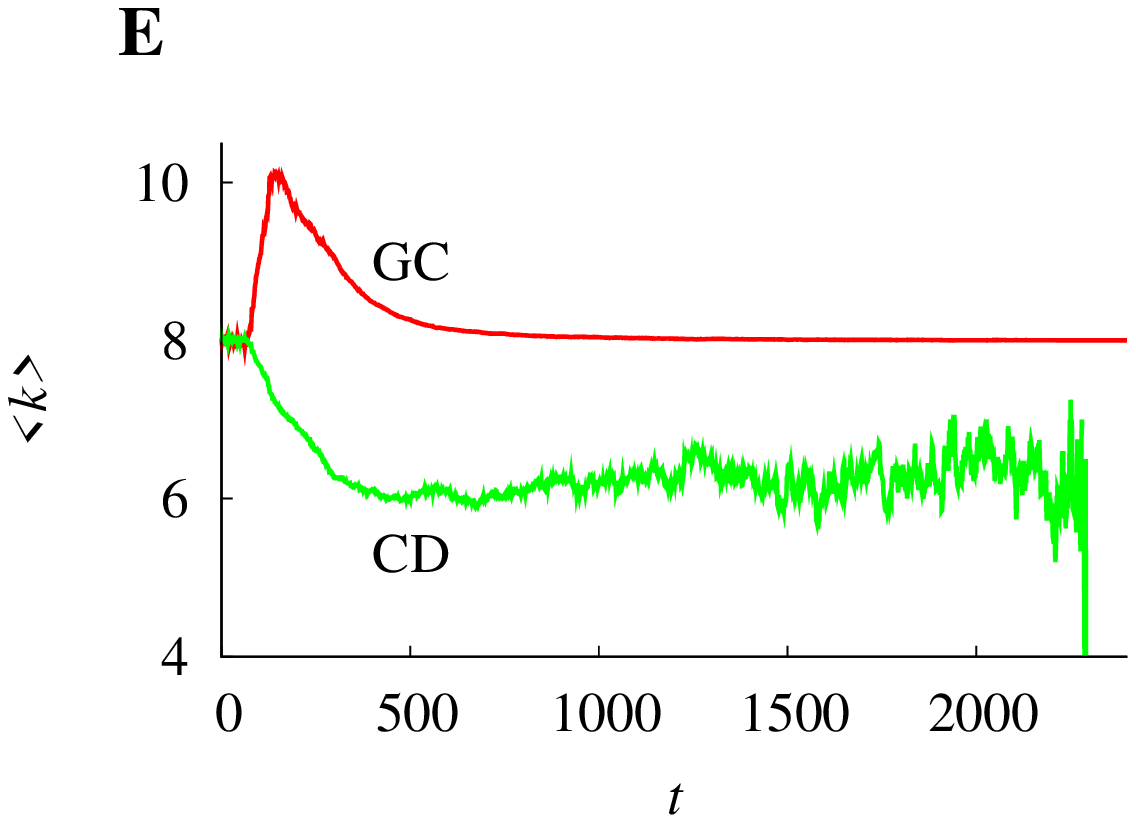}
\includegraphics[height=6cm,width=8cm]{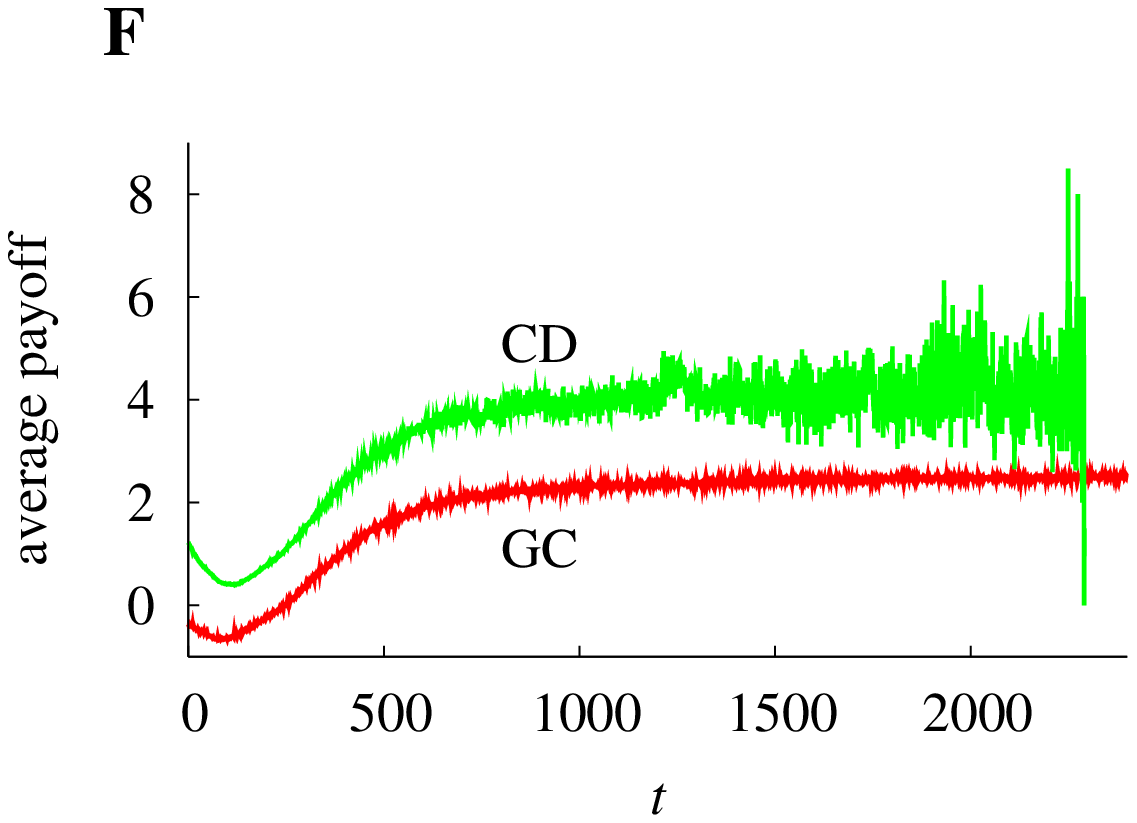}
\caption{Payoff per round for each strategy as a function of
degree at (A) $t=0$, (B) $t=200$, (C) $t=800$, and (D)
$t=2400$.  (E) Time course of
mean degree for GC and CD.  (F)
Time course of average payoff for GC and CD.
We set $\left<k\right>=8$, $p_i=0.8$, and $N_{\rm u}=200$.}
\label{fig:GC CD detail}
\end{center}
\end{figure}

To verify this prediction, we carry out Monte Carlo
simulations of the game of upstream reciprocity
on the scale-free network with a random
mixture of GCs and CDs. We set $b=1.5$.
The probability that each player is initially GC or CD is
0.5.
Figure~\ref{fig:GC CD detail}A shows the dependence of the
payoff per round on the degree of the player,
just before the first update (\textit{i.e.}, $t=0$).
Each data point corresponds to the payoff per round averaged over all
players having the same degree and same strategy. 
For each strategy, the payoff per round is roughly proportional to the
degree. CDs generally 
gain larger payoffs than GCs, because CDs exploit GCs
in the neighborhood. 

However, from \FIG\ref{fig:GC CD detail}A, it cannot be concluded
that CD takes over GC in the evolutionary time course.  
The same statistics are plotted at $t=200$ in 
\FIG\ref{fig:GC CD detail}B. As in the case of
\FIG\ref{fig:GC CD detail}A,
CD gains more than GC at the same degree.
At this stage, however,
most hubs are occupied by GCs for the following reason.
There are usually some GCs in the
neighborhood of a GC hub, which is also the case under
random initial condition. Then,
the GC hub tends to gain a large payoff because GC neighbors
help the GC hub. As a result of evolution,
GC will spread from the hub to the
neighbors, which further increases the payoff of the GC hub.
Suppose a situation where
CDs invade neighbors of the GC hub and exploit it.
Because the degrees of these CDs are generally not large,
the CDs
cannot be helped by many players even if the neighborhood is occupied
by GCs. Therefore,
the CDs would not gain the
payoff per round as large as that of the GC hub.
Accordingly, GC tends to be stabilized at the hub.
In contrast, if CD spreads from the hub
to the neighbors, the CD hub will obtain a small payoff.
Then, a GC in the
neighborhood of the CD hub may take over the hub;
CDs occupying hubs are not stabilized.
GCs gradually spread from hubs to players having small
degrees (\FIG\ref{fig:GC CD detail}C), and the entire network is
eventually occupied by GCs after sufficient rounds
(\FIG\ref{fig:GC CD detail}D).

The time courses of the mean degree of GCs and that of CDs
corresponding to the run shown in \FIG\ref{fig:GC CD detail}A--D are
plotted in \FIG\ref{fig:GC CD detail}E. First, the mean degree of GCs
grows until most hubs are occupied by the GCs. It then relaxes to
$\left<k\right>=8$.  The mean degree of the CDs is considerably
smaller than $\left<k\right>=8$ throughout the run.

The time courses of the average payoff per round
of GCs and that of CDs, corresponding to the same run as above,
are shown in \FIG\ref{fig:GC CD detail}F. Initially, the two
average payoffs decrease because CDs replace GCs.
Then, GCs are stabilized at hubs, and the GCs
begin to disseminate to increase
the average payoff of both GCs and CDs.
At any $t$, CDs earn more than GCs on an
average. However, this does not imply that CD invades GC macroscopically.
As shown in
\FIG\ref{fig:GC CD detail}B--C, the players with the 
largest payoffs are GC hubs rather than CDs. 
A player is chosen as a potential parent to be mimicked by other players
with the probability proportional 
to its degree \cite{Newman03siam,Noh04prl}.
In the scale-free network,
a neighbor of an arbitrary player tends to be a hub, and then
GC hubs are imitated by relatively many players.
Therefore,
while the average payoff of CDs is maintained at a larger
value than that of GCs, the fraction of CDs gradually
decreases until the CD becomes extinct.
The relative strength of a strategy
in reproduction is determined not by the average payoff of the players
using that strategy
but by the degree-weighted average payoff of these players.

\begin{figure}
\begin{center}
\includegraphics[height=6cm,width=8cm]{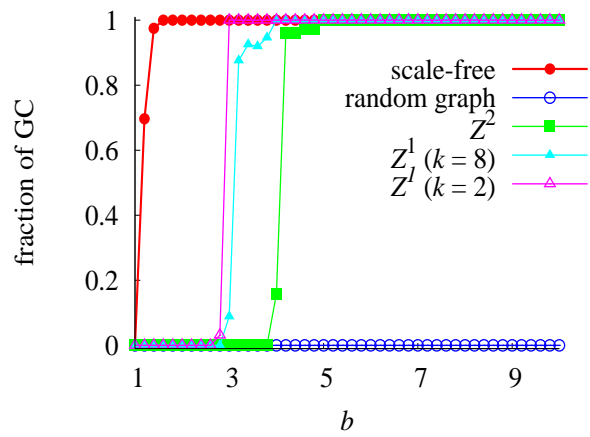}
\caption{Final fractions of GC in
various networks when players initially adopt either GC or CD.
We set $\left<k\right>=8$, $p_i=0.8$, and $N_{\rm u}=200$.}
\label{fig:GC CD}
\end{center}
\end{figure}

The scenario of evolution of helping behavior described above requires
heterogeneous degree distribution.
To compare different networks, at a given value of $b$, we 
generate five realizations of the network and 
carry out 10 runs on each network 
for the scale-free network and the regular random graph, which are
generated from stochastic algorithms. For the other three deterministic
networks, we carry out 50 runs on the network.
The average of the
final fraction of GC, obtained from the 50 runs,
is plotted against $b$ in
\FIG\ref{fig:GC CD}. In all the networks, except the regular random
graph, the fraction of GC increases with $b$. In fact, the fraction jumps
from unanimity of CD to that of GC at a threshold value
of $b$. The threshold value of $b$
above which GCs survive is considerably smaller in the
scale-free network than in the other networks. Heterogeneous
networks promote the evolution of helping behavior. Among the other
networks, the threshold value of $b$
is the smallest in the cycle. The next
smallest value is the extended cycle and then the square
lattice. The threshold value of $b$ in 
the random graph is greater than the upper limit shown in
\FIG\ref{fig:GC CD} (\textit{i.e.}, $b=10$).
Unlike the Barab\'{a}si--Albert scale-free network and
the regular random graph, the other three networks, \textit{i.e.},
the cycle, the
extended cycle, and the square lattice, are capable of spatial
reciprocity. This fact explains why these three
networks accommodate more GCs as 
compared to the regular random graph. However, the effect of
spatial reciprocity is smaller than the effect of the scale-free
networks, at least under the present parameter regime.

\begin{figure}
\begin{center}
\includegraphics[height=6cm,width=8cm]{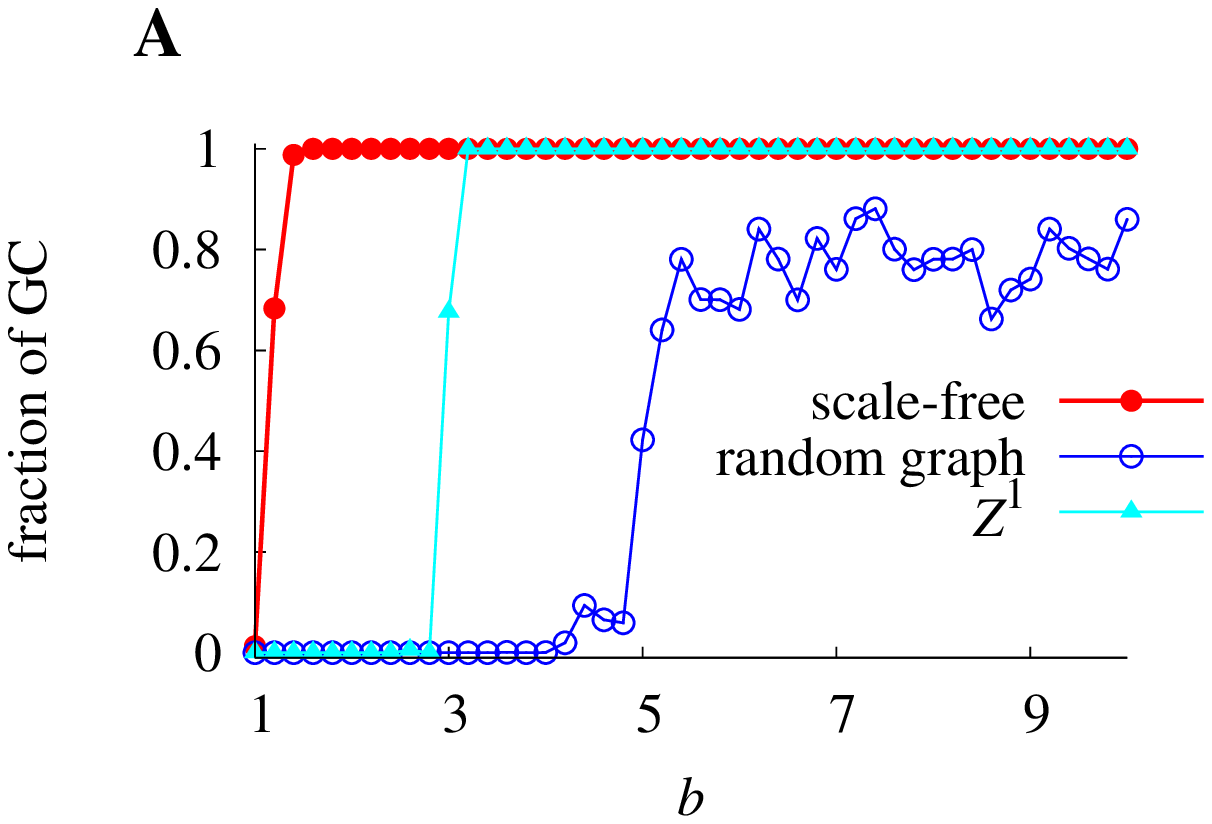}
\includegraphics[height=6cm,width=8cm]{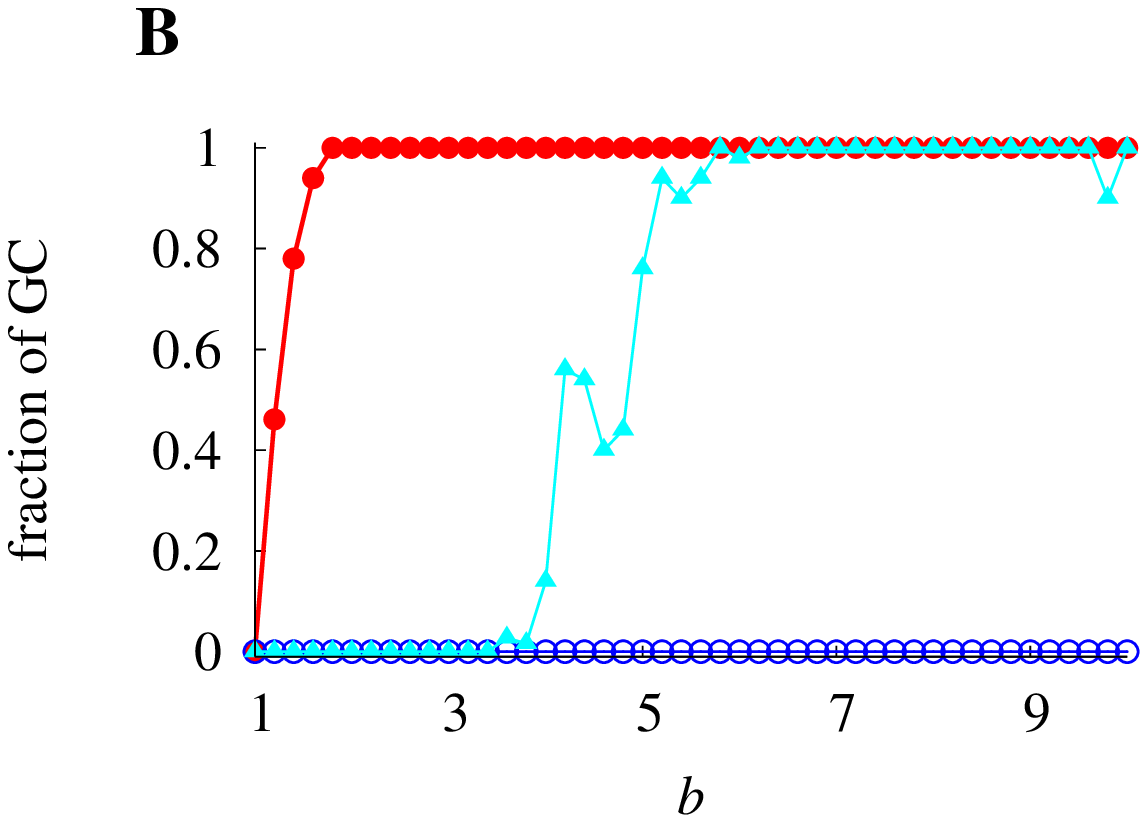}
\includegraphics[height=6cm,width=8cm]{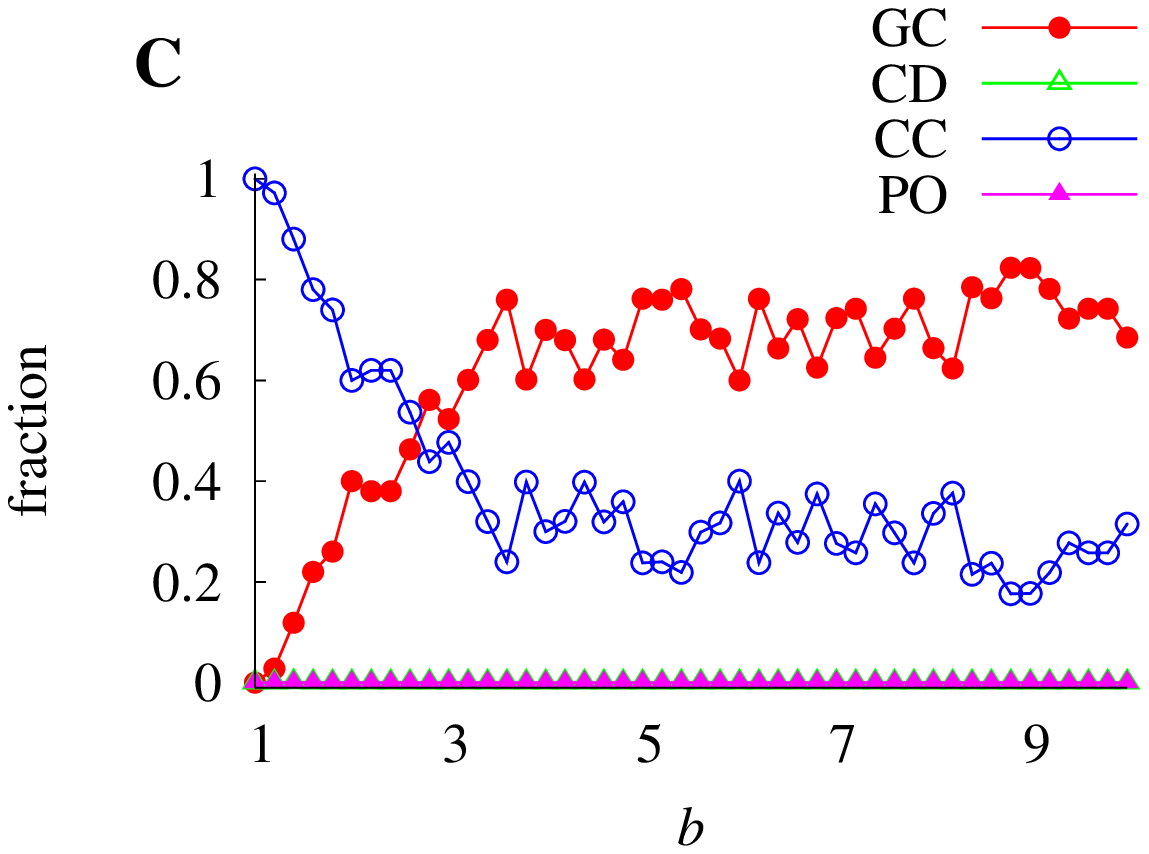}
\includegraphics[height=6cm,width=8cm]{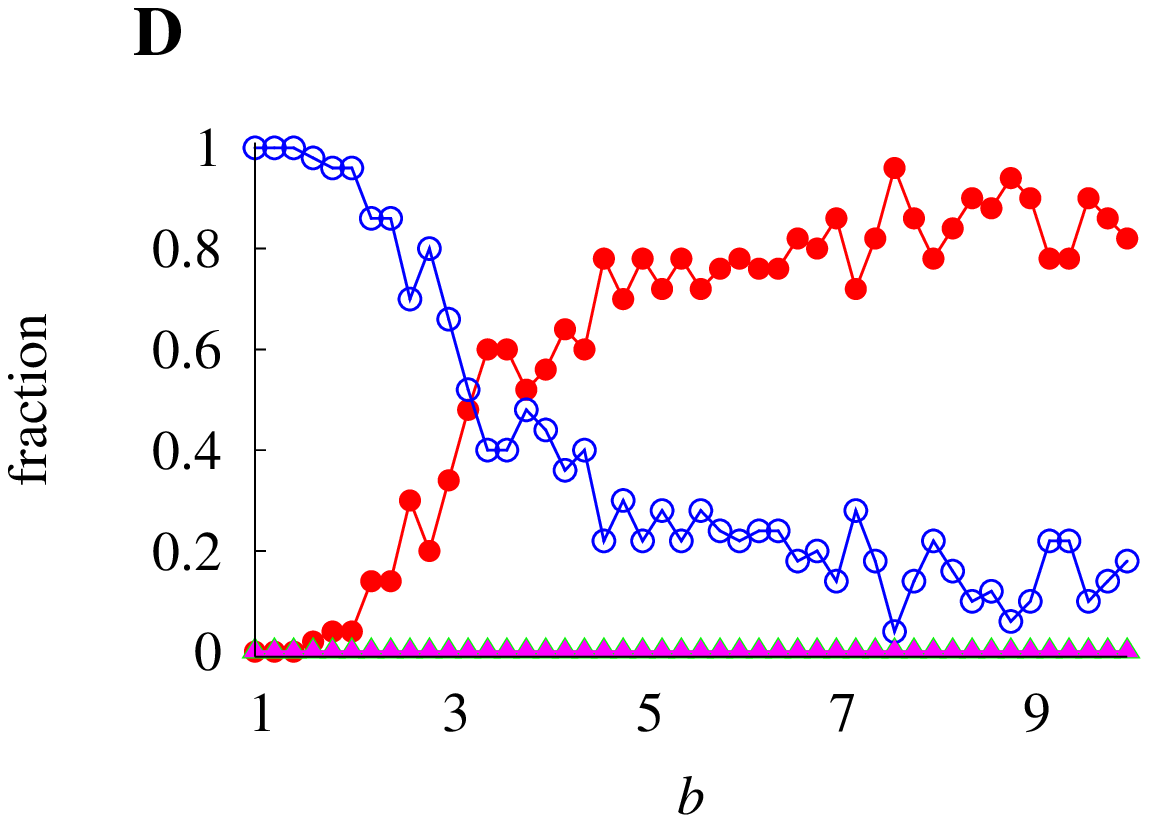}
\caption{Results for different values of $\left<k\right>$.
(A, B) Final fractions of GC in various networks
when players initially adopt either GC or CD.
We set (A) $\left<k\right>=6$ and (B) $\left<k\right>=14$.
(C, D) Final fractions of four strategies
when players initially adopt either GC, CD, CC, or PO in
the scale-free network. We set (C) $\left<k\right>=6$
and (D) $\left<k\right>=14$.}
\label{fig:different k}
\end{center}
\end{figure}

\begin{figure}
\begin{center}
\includegraphics[height=6cm,width=8cm]{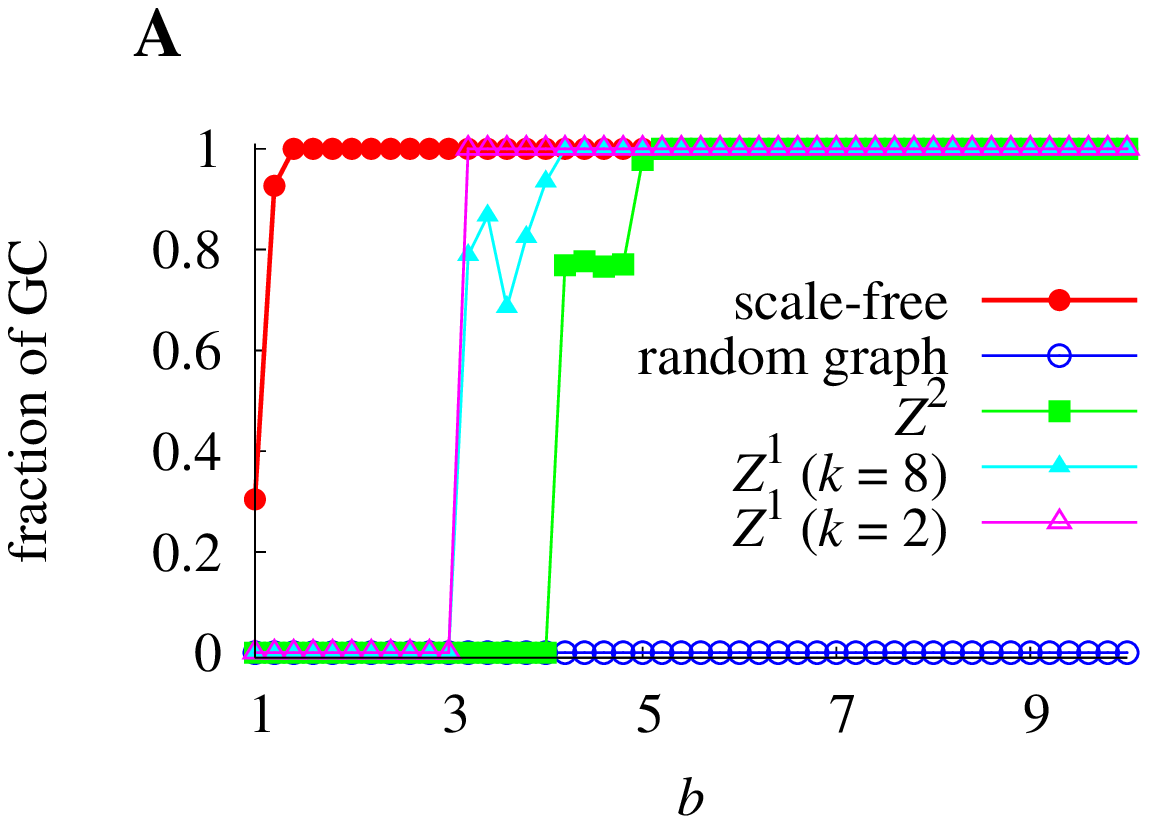}
\includegraphics[height=6cm,width=8cm]{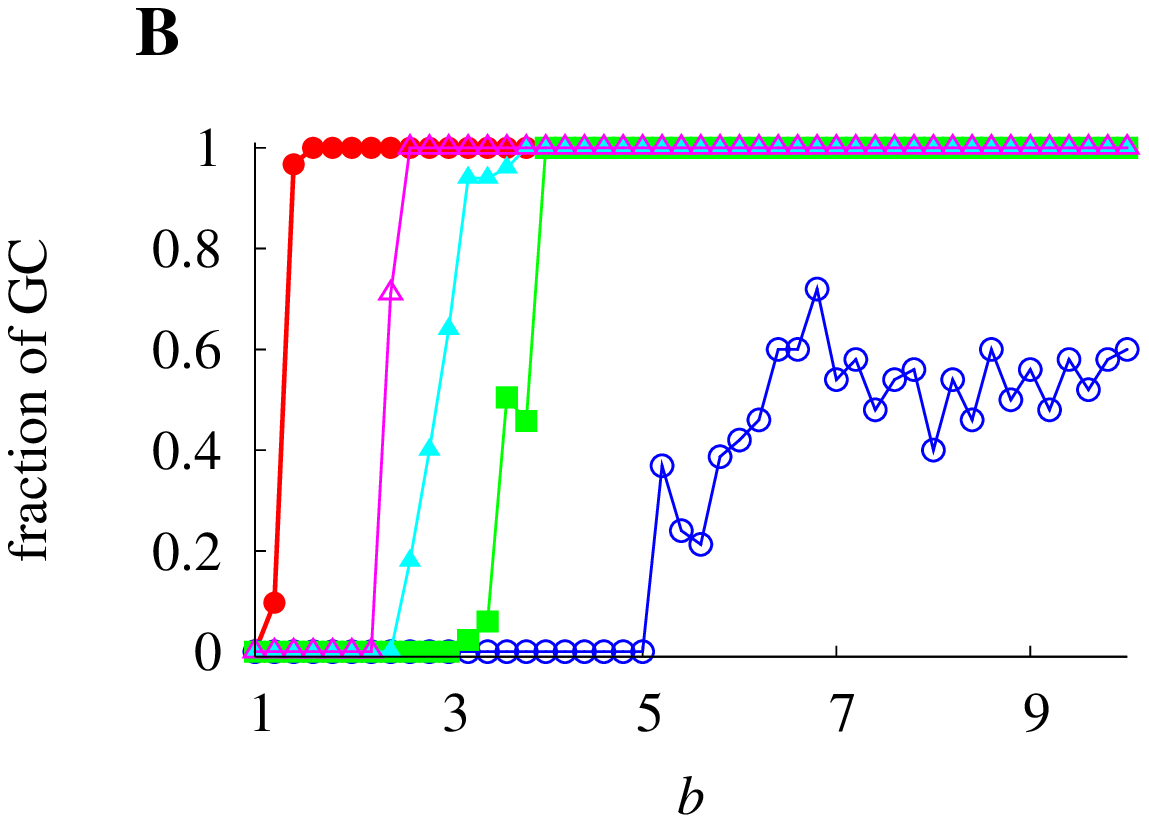}
\includegraphics[height=6cm,width=8cm]{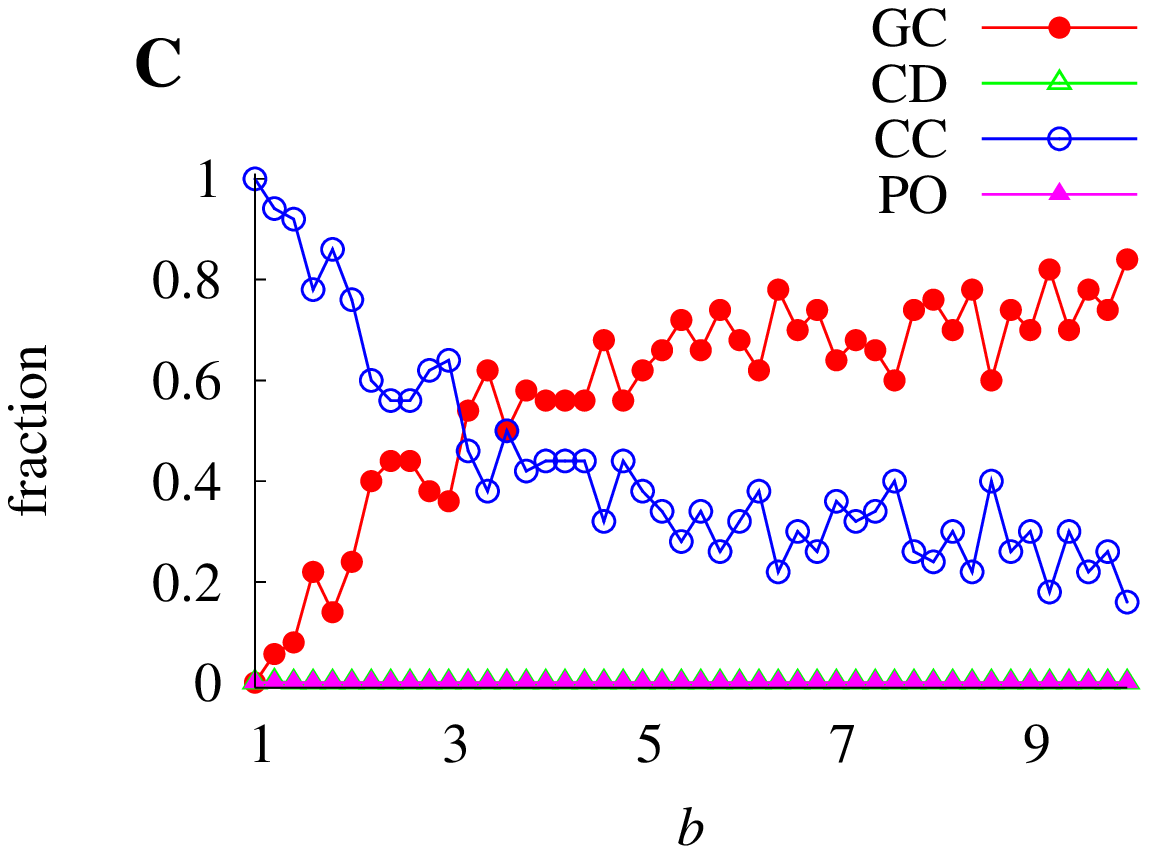}
\includegraphics[height=6cm,width=8cm]{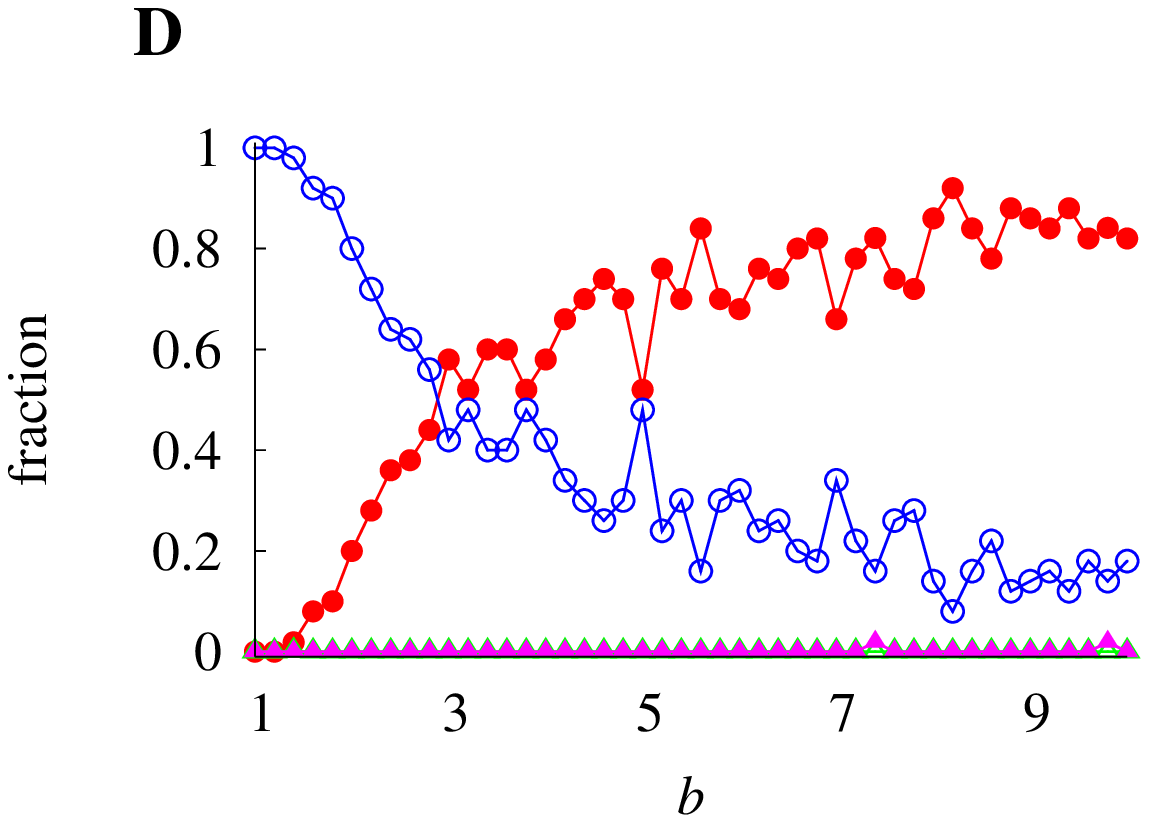}
\caption{Results for different values of $p_i$.
(A, B) Final fractions of GC in various networks
when players initially adopt either GC or CD.
We set $p_i$ for GC to (A) 0.7 and (B) 0.9.
(C, D) Final fractions of four strategies
when players initially adopt either GC, CD, CC, or PO in
the scale-free network. We set $p_i$ for GC and PO to
(C) 0.7 and (D) 0.9.}
\label{fig:different p}
\end{center}
\end{figure}

\begin{figure}
\begin{center}
\includegraphics[height=6cm,width=8cm]{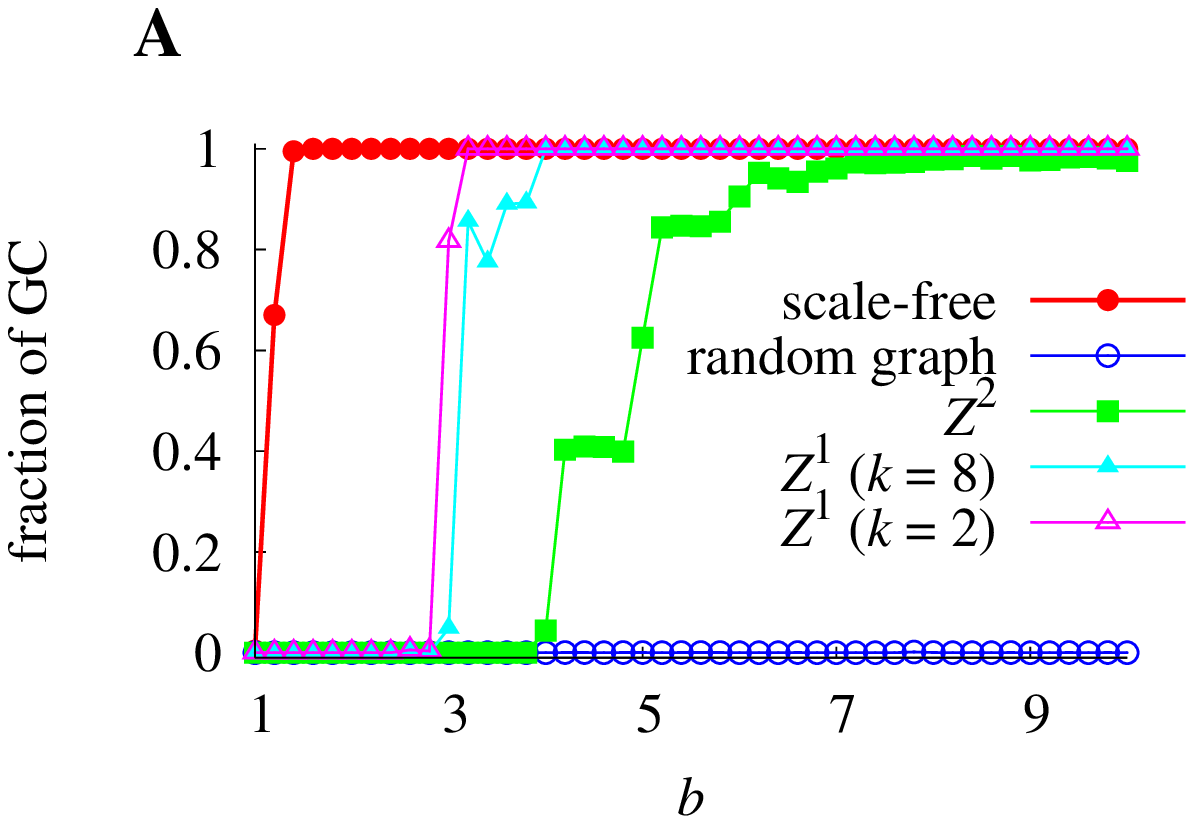}
\includegraphics[height=6cm,width=8cm]{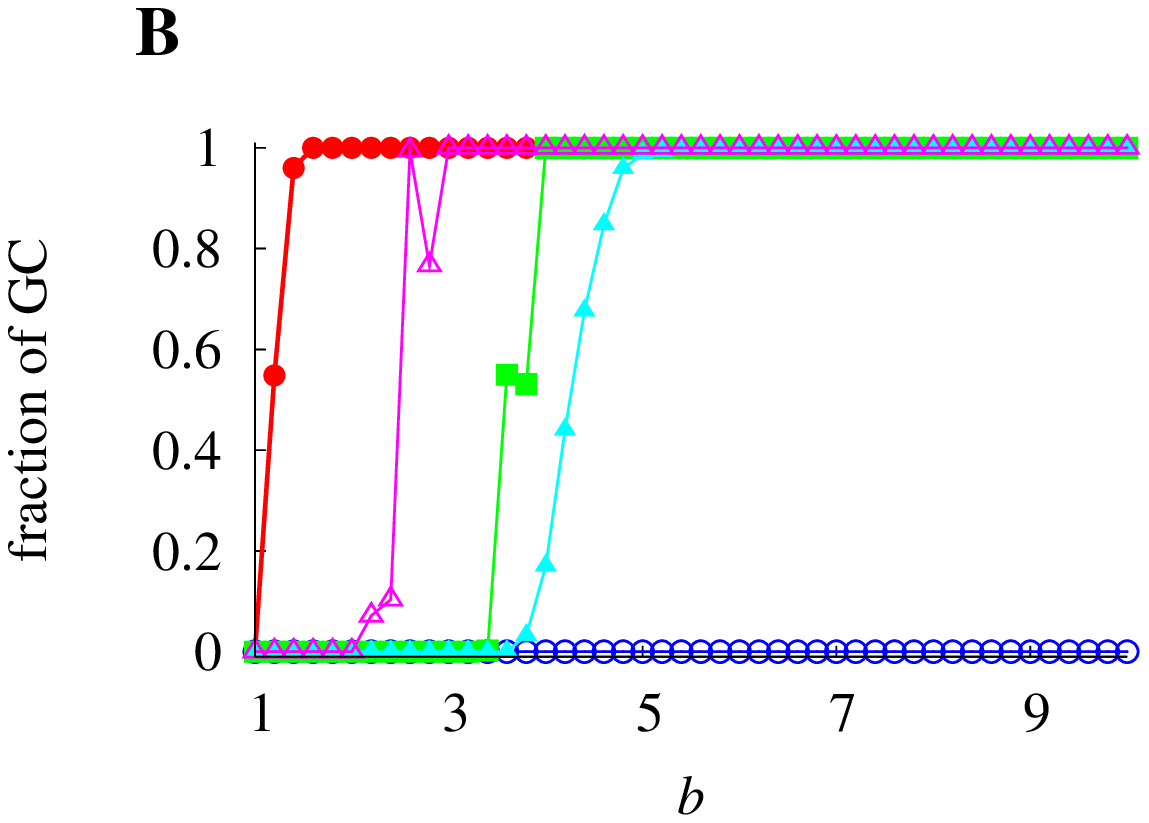}
\includegraphics[height=6cm,width=8cm]{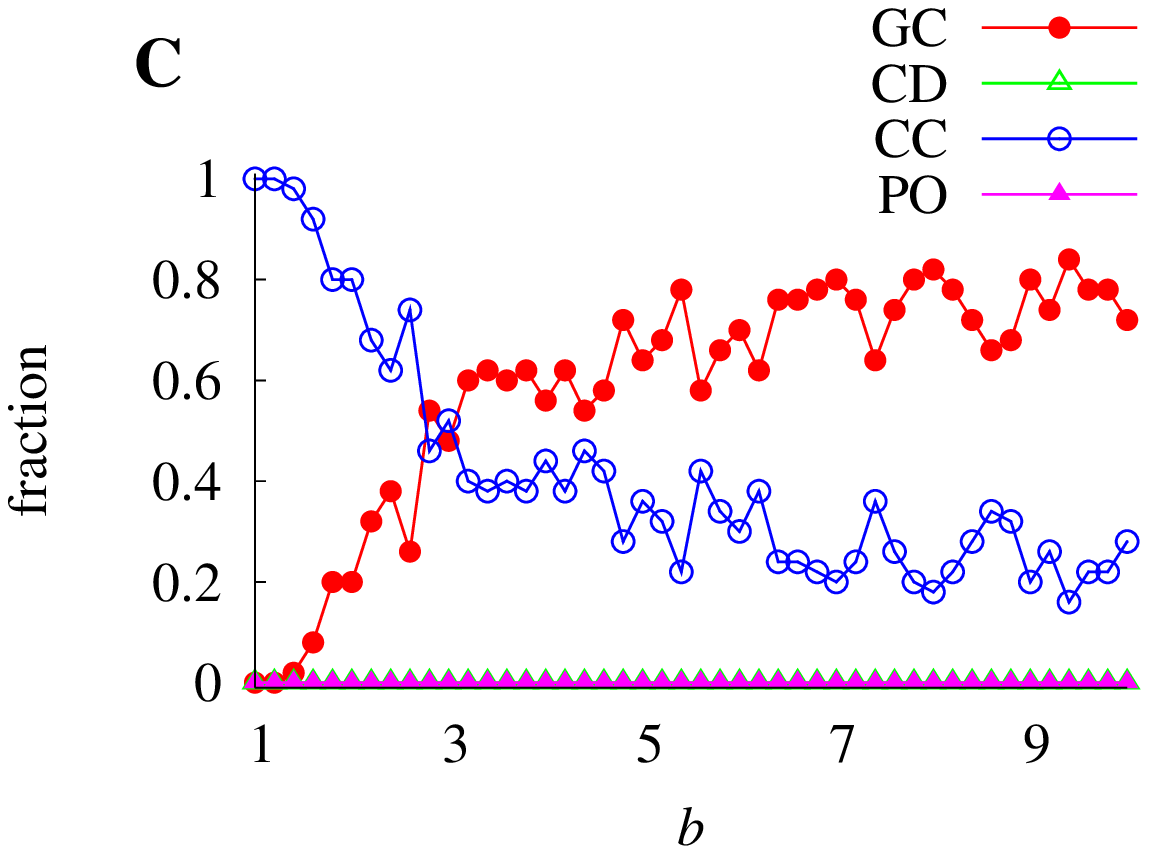}
\includegraphics[height=6cm,width=8cm]{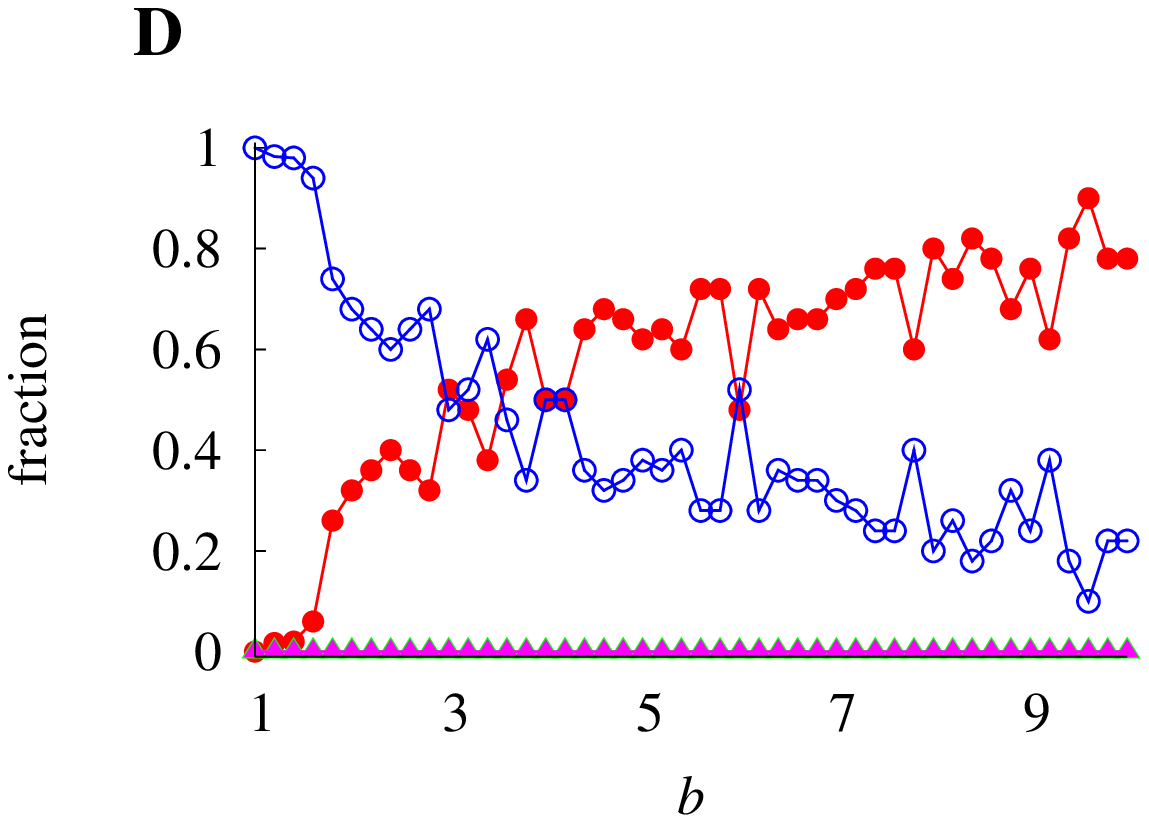}
\caption{Results for different numbers of players updated per round.
(A, B) Final fractions of GC in various networks
when players initially adopt either GC or CD.
We set (A) $N_{\rm u}=20$ and (B) $N_{\rm u}=2000$.
(C, D) Final fractions of four strategies
when players initially adopt either GC, CD, CC, or PO in
the scale-free network. We set (C) $N_{\rm u}=20$ and 
(D) $N_{\rm u}=2000$.}
\label{fig:different N_u}
\end{center}
\end{figure}

\begin{figure}
\begin{center}
\includegraphics[height=6cm,width=8cm]{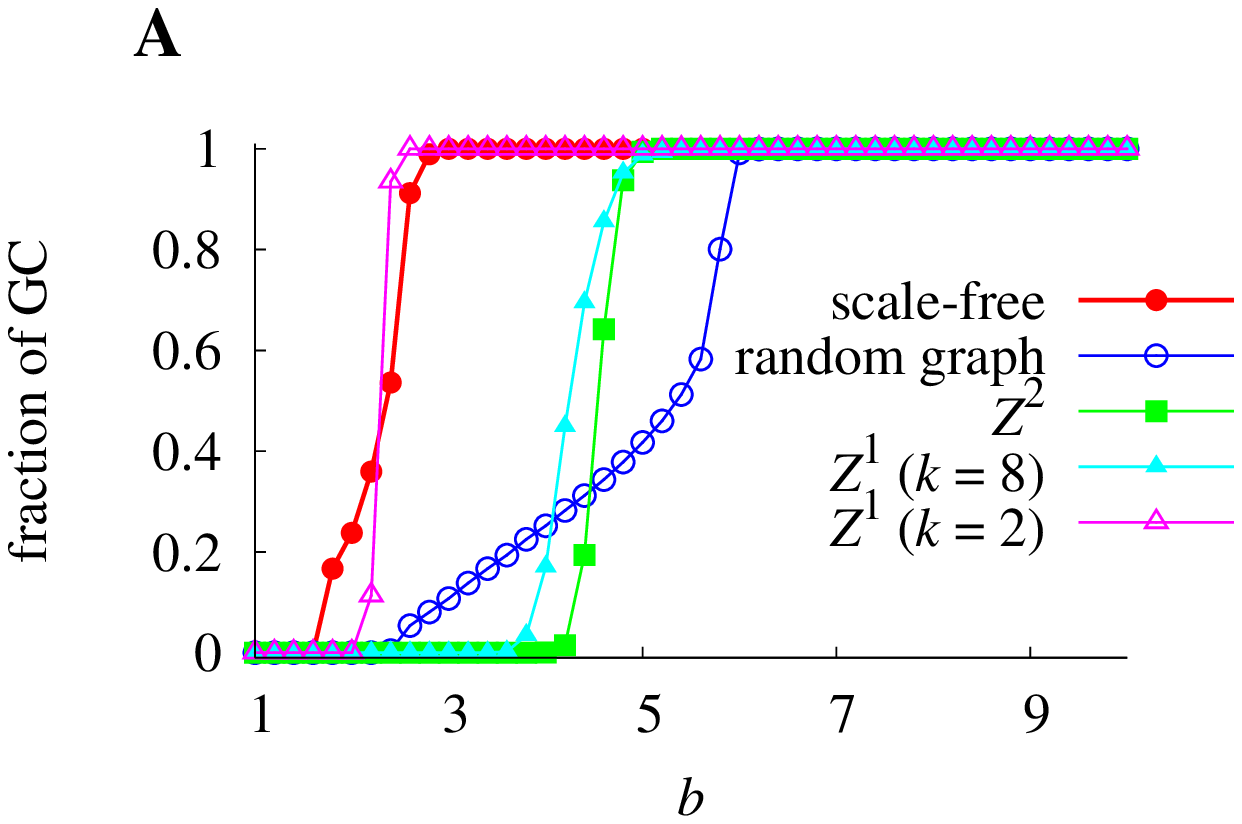}
\includegraphics[height=6cm,width=8cm]{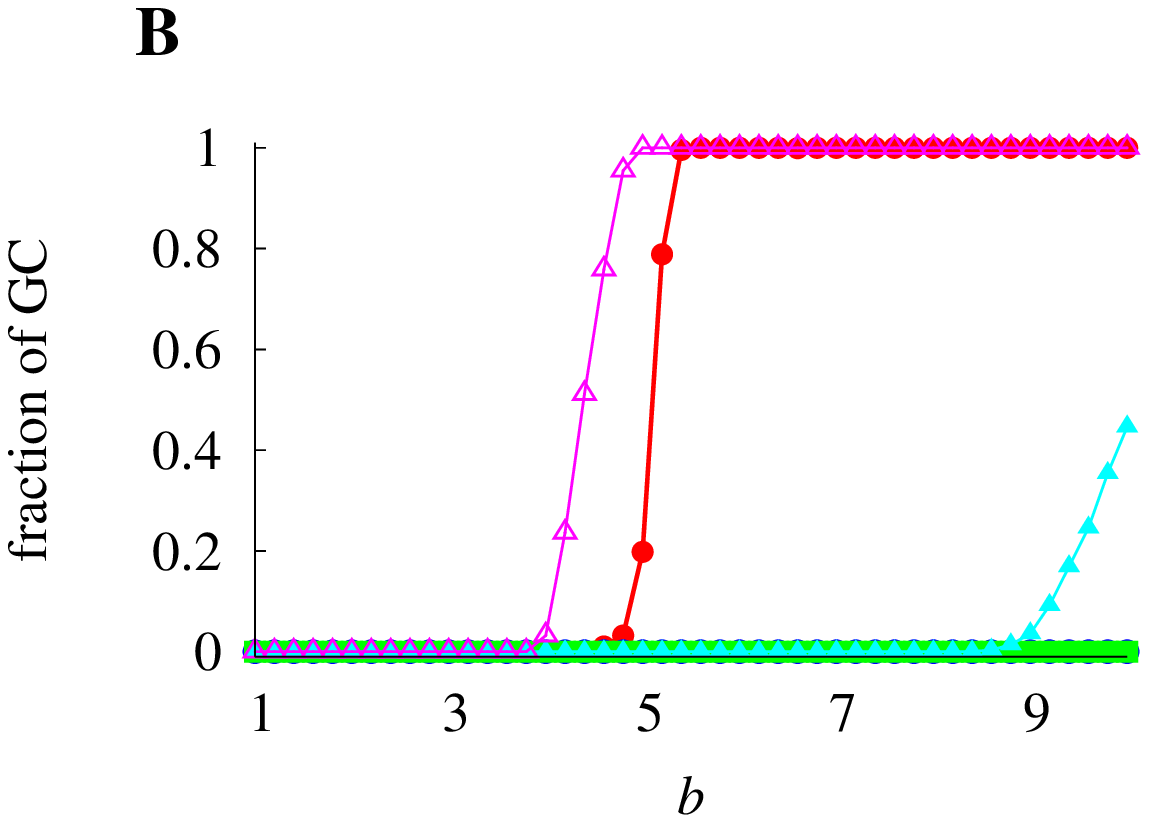}
\includegraphics[height=6cm,width=8cm]{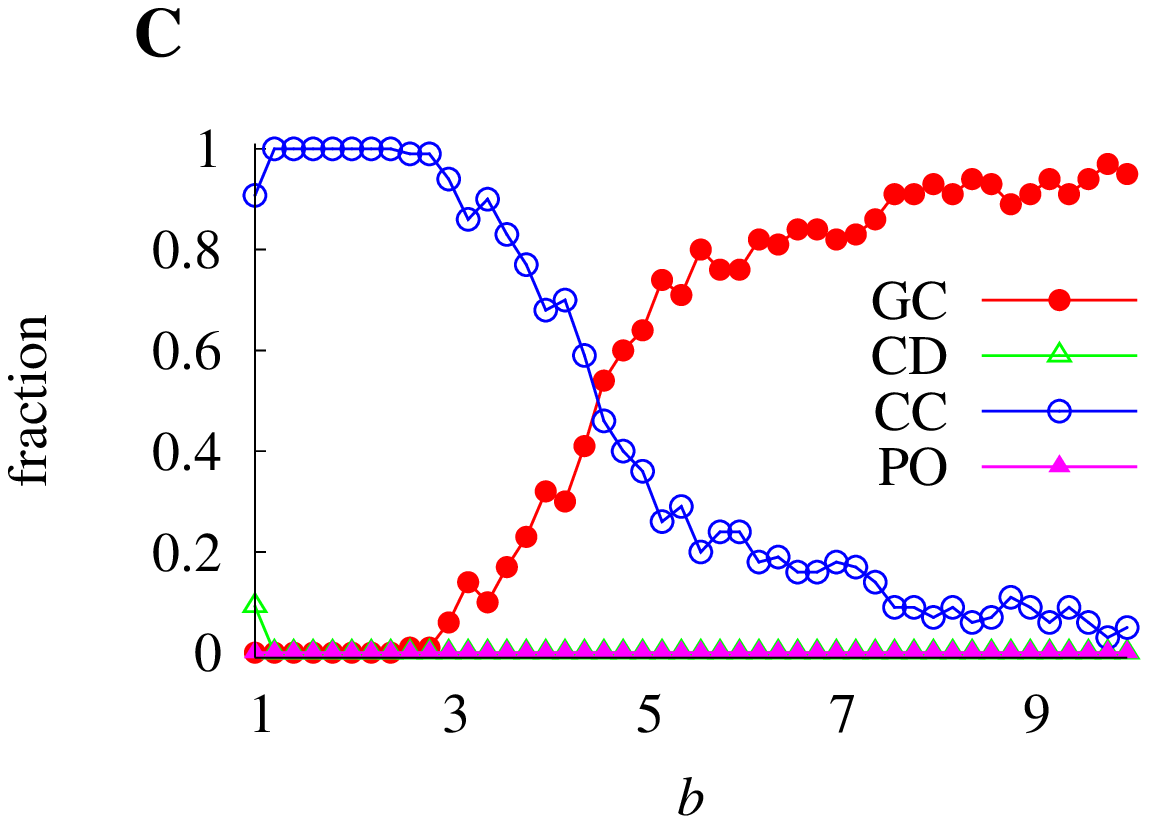}
\includegraphics[height=6cm,width=8cm]{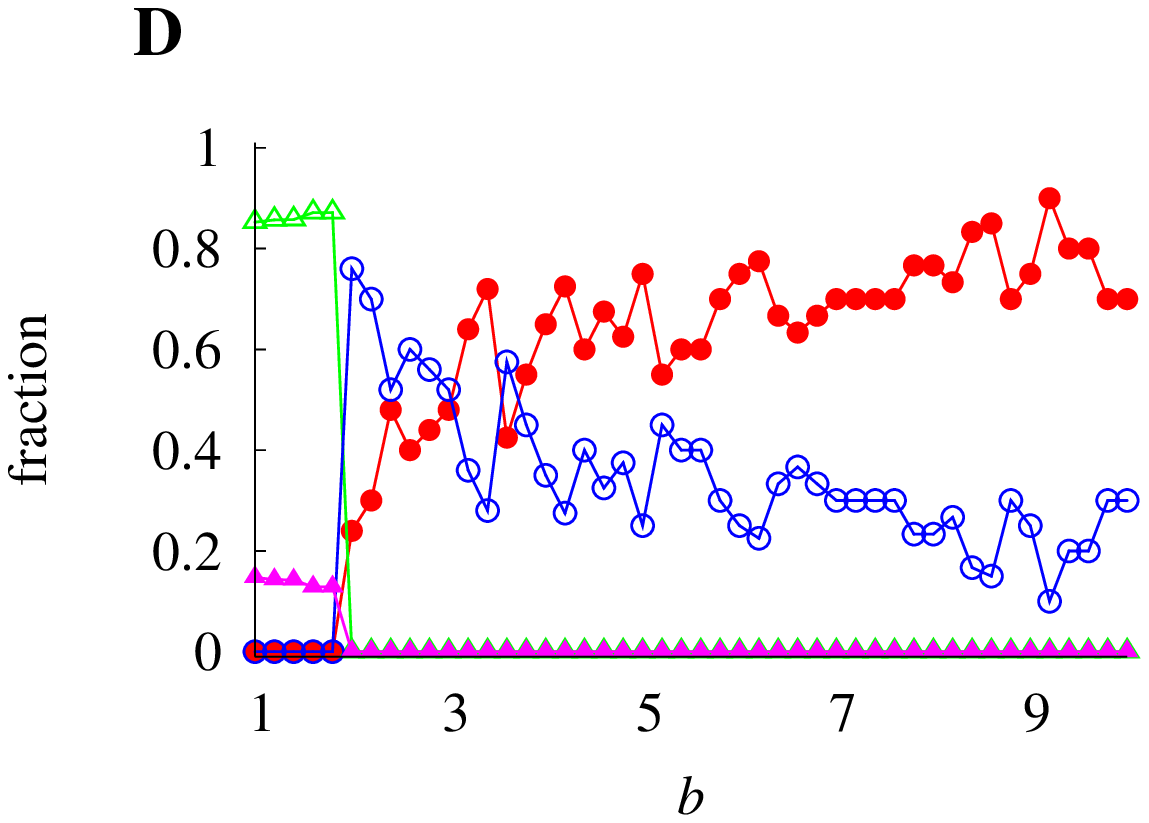}
\caption{Results for different update rules.
(A, B) Final fractions of GC in various networks
when players initially adopt either GC or CD.
We use (A) imitation update rule and (B) Fermi update rule.
(C, D) Final fractions of four strategies
when players initially adopt either GC, CD, CC, or PO in
the scale-free network. We use (C) imitation update
rule and (D) Fermi update rule.}
\label{fig:different update rules}
\end{center}
\end{figure}

We confirm that the results are qualitatively the same
for some variations of the model.
First, we change the mean degree to
$\left<k\right>=6$ (\FIG\ref{fig:different k}A)
and $\left<k\right>=14$ (\FIG\ref{fig:different k}B). 
The results are qualitatively the same as those for $\left<k\right>=8$. 
Quantitatively,
GC survives more easily for a smaller $\left<k\right>$, which coincides
with the results for the Prisoner's Dilemma on regular random
graph \cite{Ohtsuki06nat}.
Second, we change $p_i$ for GC and PO to 0.7 
(\FIG\ref{fig:different p}A)
and 0.9 (\FIG\ref{fig:different p}B).
The results are qualitatively the same as those for $p_i= 0.8$.
Quantitatively, a larger value of $p_i$ yields a larger fraction of GC.
Third, we change the number of players updated in one round to
$N_{\rm u}=20$ (\FIG\ref{fig:different N_u}A) 
and $N_{\rm u}=2000$ (\FIG\ref{fig:different N_u}B).
The results are qualitatively the same as those for $N_{\rm u}=200$.
Fourth, we show the effect of different stochastic update rules.
In the imitation rule \cite{Ohtsuki06nat},
at the end of each round,
potentially updated player $v_i$ 
selects a potential parent out of the $k_i+1$ players,
\textit{i.e.}, $v_i$ and the $k_i$ neighbors of $v_i$. The probability
that a node is selected as the parent is proportional to the
payoff. When the payoff is negative, we set this probability to zero.
In the Fermi rule
(\textit{e.g.}, Szab\'{o} and T\H{o}ke, 1998; Traulsen et al., 2006),
$v_i$ selects a potential parent $v_j$ 
out of the $k_i$ neighbors with equal
probability and copies the strategy of $v_j$ with probability
$\left[1+\exp(\beta((\mbox{payoff of player } v_i) - (\mbox{payoff
of player } v_j)))\right]^{-1}$.
Otherwise, $v_j$ copies the stragegy of $v_i$.
The results for the imitation rule
and those for the Fermi rule with $\beta=0.2$
are shown in \FIGS\ref{fig:different update rules}A and 
\ref{fig:different update rules}B, respectively.
The results resemble those for 
the deterministic update rule. Although
the one-dimensional chain allows for GC at small values of $b$,
as comparable or even smaller than the values for the scale-free
network, our main result that heterogeneous
networks enhances generous cooperators as compared to homogeneous
networks is not violated.

In the case of the cycle, the threshold value of $b$ above which the
GC survives the invasion by CD has been obtained 
for a different update rule in the limit of weak
selection \cite{Nowak07}.  The survival of the GC is possible when
$b/c > f(p)$, where $f(p) =$ $\left[8+2p+8\sqrt{1-p^{2}}\right]\big/$
$\left[3+4p+\sqrt{1-p^{2}}\right]$.  Because we set $c=1$ and $p=0.8$,
the theoretical threshold in this case is equal to $f(0.8) =
2.12$. Figure~\ref{fig:GC CD} indicates that the GC survives when $b$
is larger than approximately 2.6 in the numerical simulations;
this value is not too far from
the theoretical value.  The discrepancy between
the theoretical and numerical results is probably attributed to the
use of different update rules (stochastic versus deterministic), the
difference in selection pressure (weak selection versus strong
selection), and/or the difference in the
boundary condition of the network (open end
versus periodic boundary condition).

\subsection{GC versus CC}\label{sub:GC CC}

Next, we examine the case in which GCs and CCs are initially present.
Although GC and CC are both cooperative in a classical sense, the
GC is more cooperative than the CC in a game of upstream
reciprocity. Similar to the case considered in \SEC\ref{sub:GC CD}, we
start each Monte Carlo simulation using an equal fraction of GCs and
CCs. In contrast to a population composed of GCs and CDs, in this
case, the unanimity of GC or that of CC, instead of a mixture of GC
and CC, is reached very often in the final round of runs in the
scale-free network and the square lattice. This unanimity is attained
even if the number of rounds is set to a small value.  If all runs end
up at unanimity, the fraction of GC is equal to the fraction of runs
in which unanimity of the GC is reached.  This quantity is discretized
by the number of runs.  Therefore, we carry out 100 runs in the
scale-free network and the square lattice to overcome the
discretization effect. In the other networks, we carry out 50 runs as
in the previous case.

\begin{figure}
\begin{center}
\includegraphics[height=6cm,width=8cm]{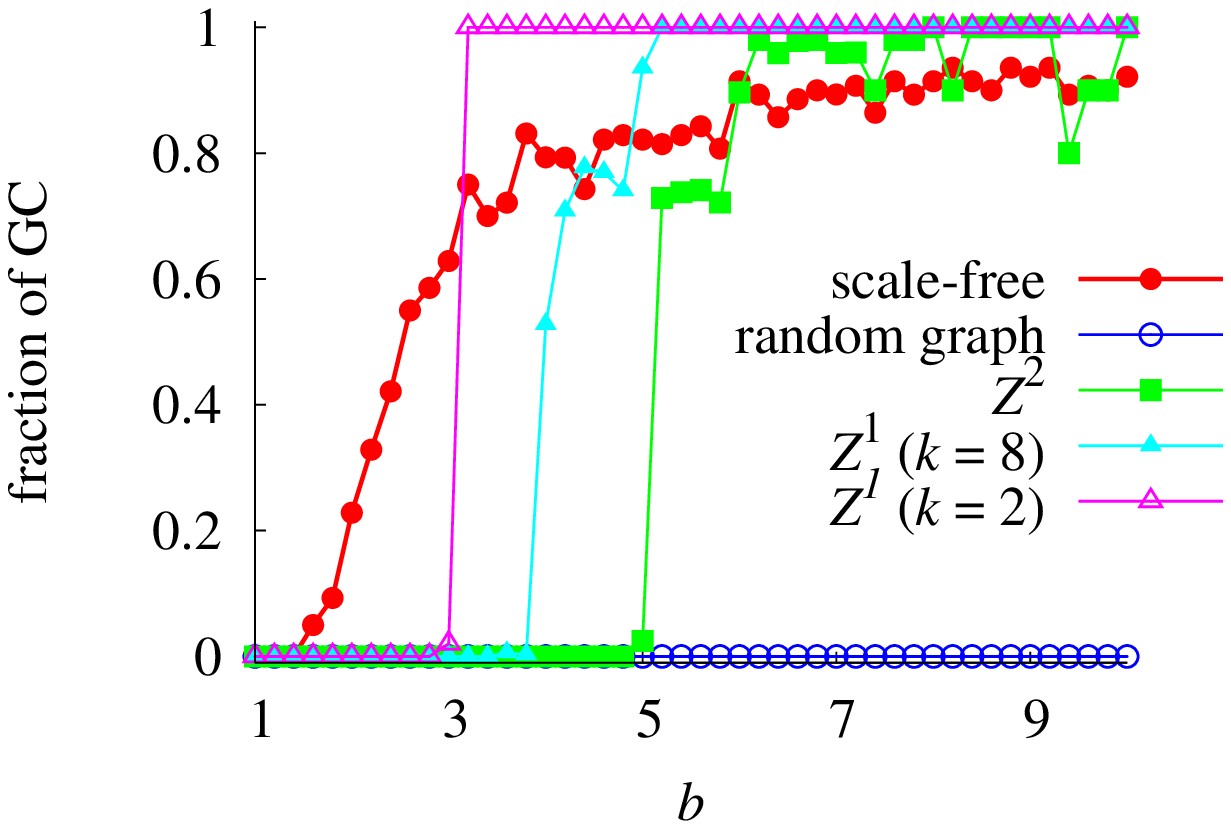}
\caption{Final fractions of GC in various networks
when players initially adopt either GC or CC.}
\label{fig:GC CC}
\end{center}
\end{figure}

The final fraction of the GC in different
networks is shown in \FIG\ref{fig:GC CC}. The scale-free network
enhances the evolution of the GC to a greater extent
than the other networks,
except at large values of $b$. This result and the ordering of
the five networks according to the threshold value of $b$ above which
the GC evolves are consistent with those obtained in the case of
the population of GCs and CDs
(\SEC\ref{sub:GC CD}).
The threshold value of $b$ in 
the random graph is greater than the upper limit shown in
\FIG\ref{fig:GC CC} (\textit{i.e.}, $b=10$).

In the case of the cycle, it has been theoretically shown 
for the original model
that the GC survives the
invasion by CC when $b/c > f(0.8) = 2.12$ \cite{Nowak07}.  In
\FIG\ref{fig:GC CC}, the GC survives in the cycle when $b/c\ge 3.0$,
which is of the same order as
the theoretically predicted value for the original model.

\subsection{GC versus PO}\label{sub:GC PO}

In this section, we investigate the population composed of GCs and
POs. Recall that, even though PO is cooperative in that
it passes on helping behavior to a neighbor, the GC is more
cooperative in comparison
because it initiates a chain of helping behavior and PO does not.

\begin{figure}
\begin{center}
\includegraphics[height=6cm,width=8cm]{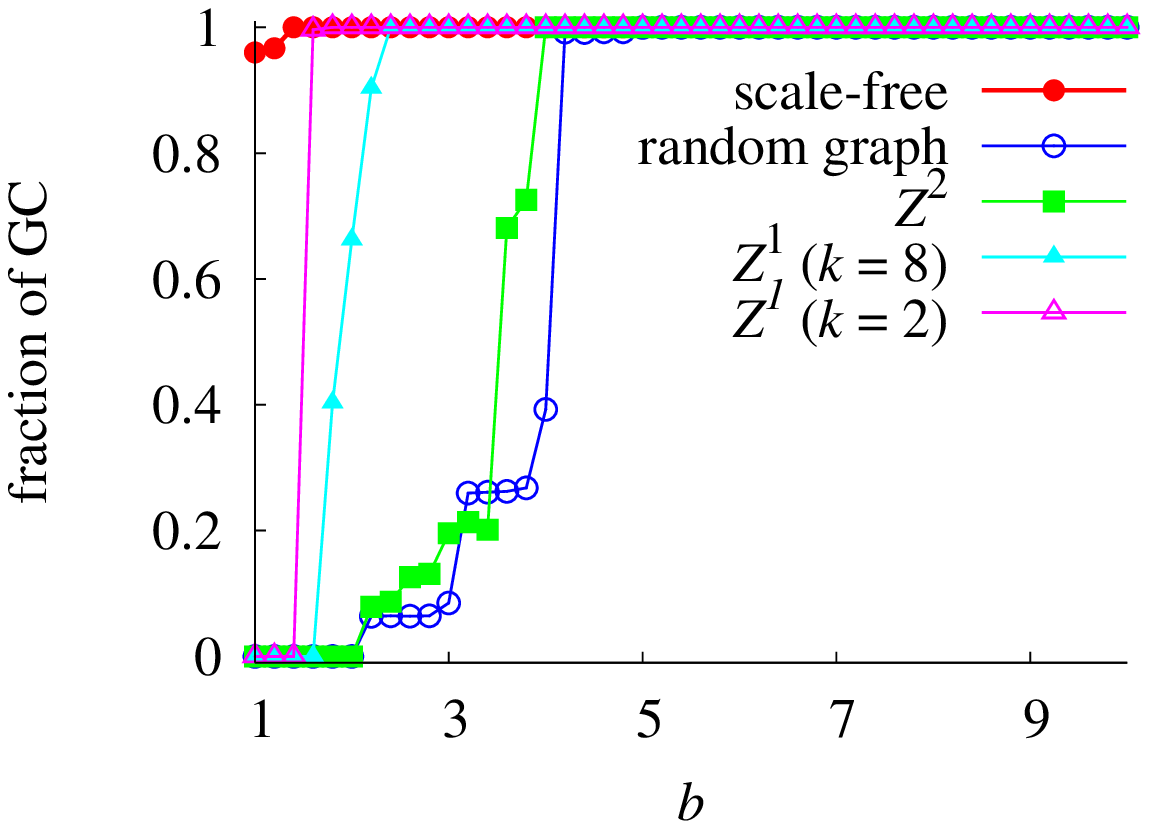}
\caption{Final fractions of GC in various networks
when players initially adopt either GC or PO.}
\label{fig:GC PO}
\end{center}
\end{figure}

The final fractions of the GC obtained from 50 runs in different networks
are compared in
\FIG\ref{fig:GC PO}. Similar to the results
reported in \SECS\ref{sub:GC CD} and \ref{sub:GC CC}, the scale-free
network yields the largest fraction of the GC.  The ordering of the five
networks according to the threshold value $b$ is also consistent with
those obtained in the population of GCs and CDs (\SEC\ref{sub:GC CD}) and 
that of GCs and CCs (\SEC\ref{sub:GC CC}).

Theoretically, GC survives for the original model
in the cycle when $b/c > g(p)$, where $g(p)
=$ $\left[p\left(3+3p+\sqrt{1-p^{2}}\right)\right]\big/$
$\left[\left(1+2p\right)\left(1+p-\sqrt{1-p^{2}}\right)\right]$
\cite{Nowak07}.  In our simulations, 
the threshold is estimated to be $g(0.8)
= 1.54$. Figure~\ref{fig:GC PO} suggests that the threshold is about
$1.5$, which is close to the theoretical value for the original model.

\subsection{Populations comprising four strategies}\label{sub:4 strategies}

We examine the dynamics of a population in which all four strategies
are initially present. Each player is assumed to adopt either strategy
independently with probability $1/4$.  Similar to the case of the
population of GCs and CCs, most runs end up at unanimity of one
strategy in the scale-free network and the extended cycle.
Therefore, we carry out 100 runs for these two networks
to enhance the precision in the computed fraction of 
different strategies.
For the other networks, we carry out 50 runs.

\begin{figure}
\begin{center}
\includegraphics[height=6cm,width=8cm]{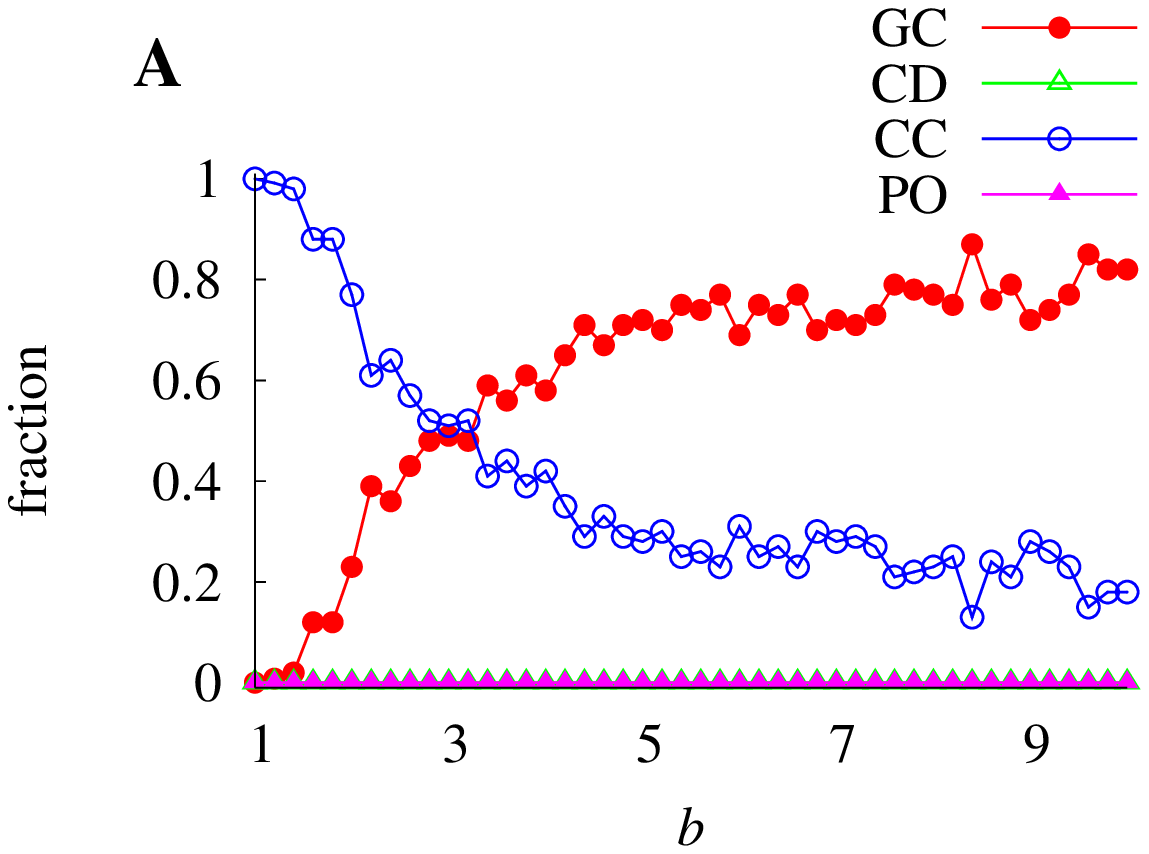}
\includegraphics[height=6cm,width=8cm]{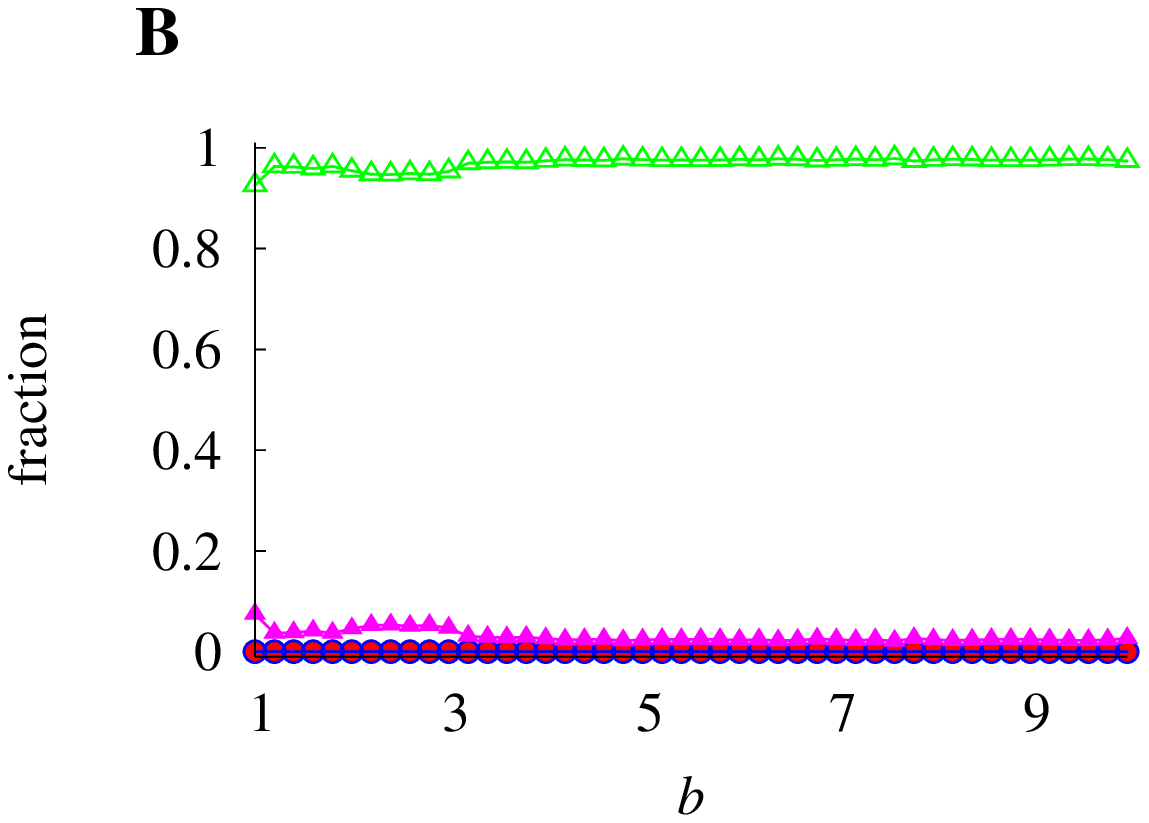}
\includegraphics[height=6cm,width=8cm]{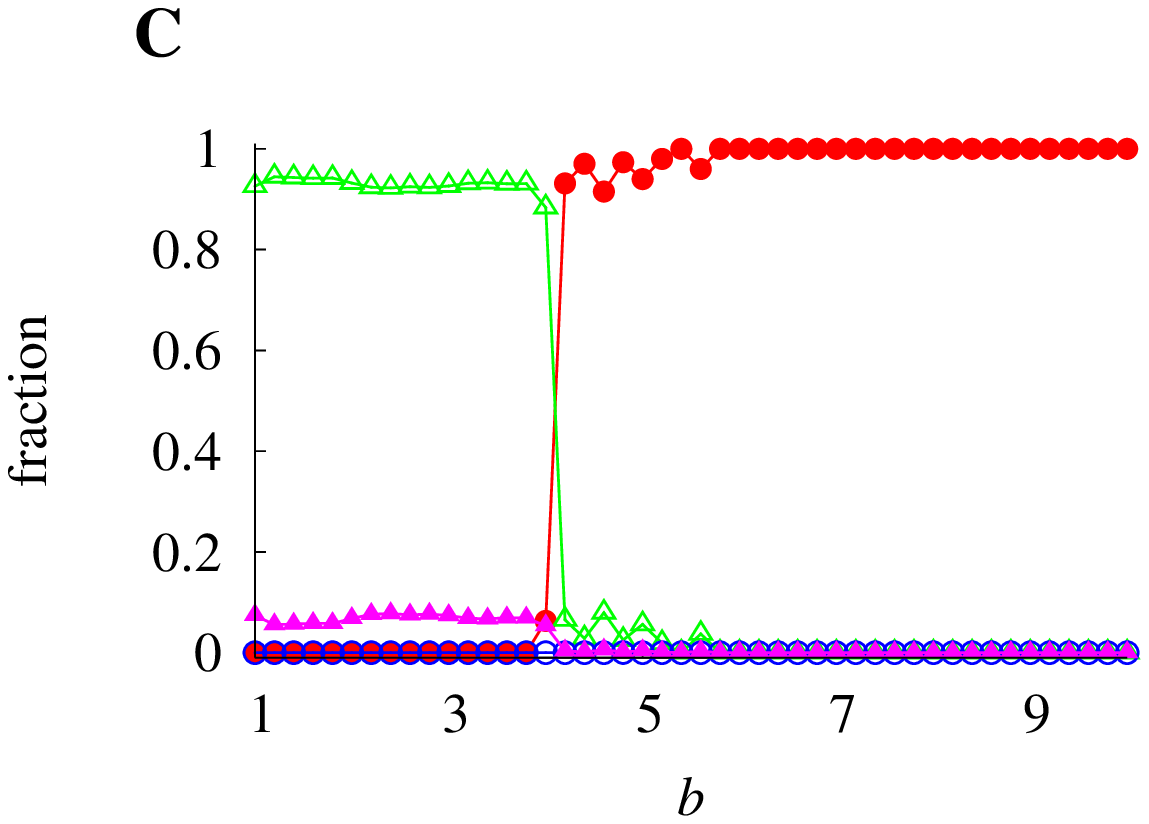}
\includegraphics[height=6cm,width=8cm]{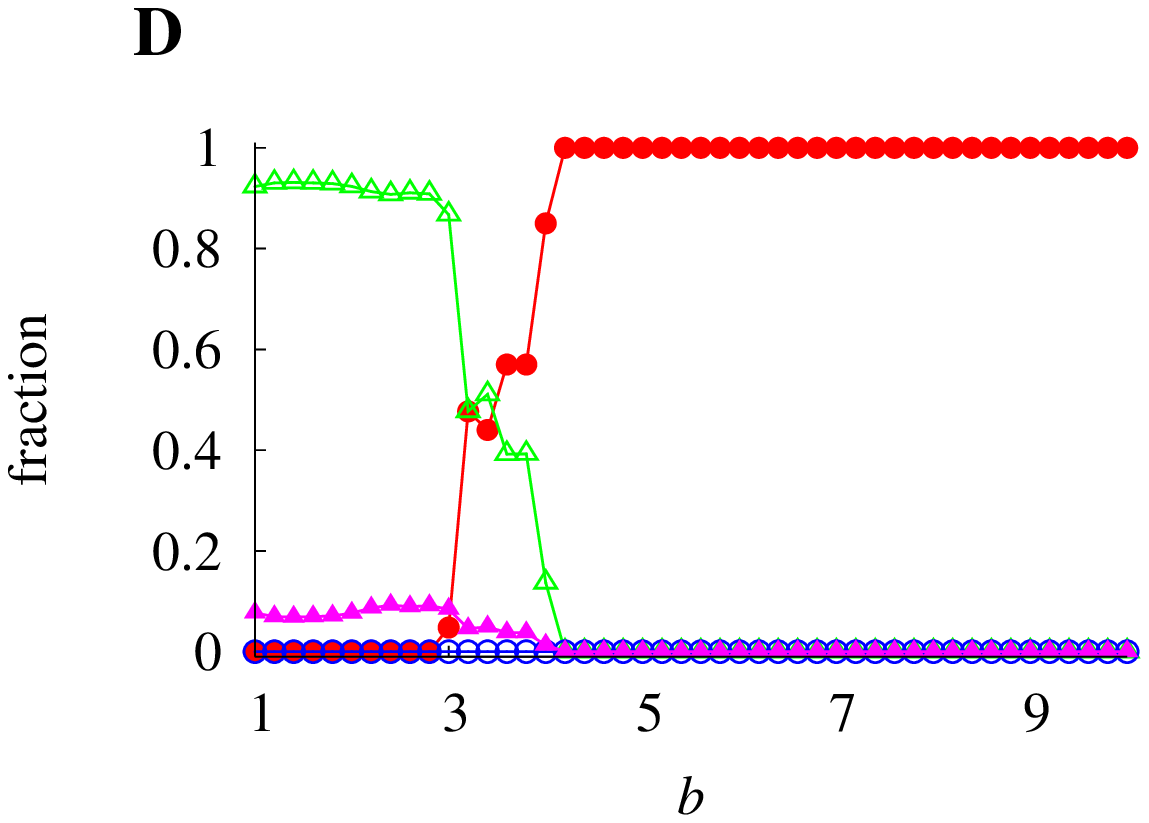}
\includegraphics[height=6cm,width=8cm]{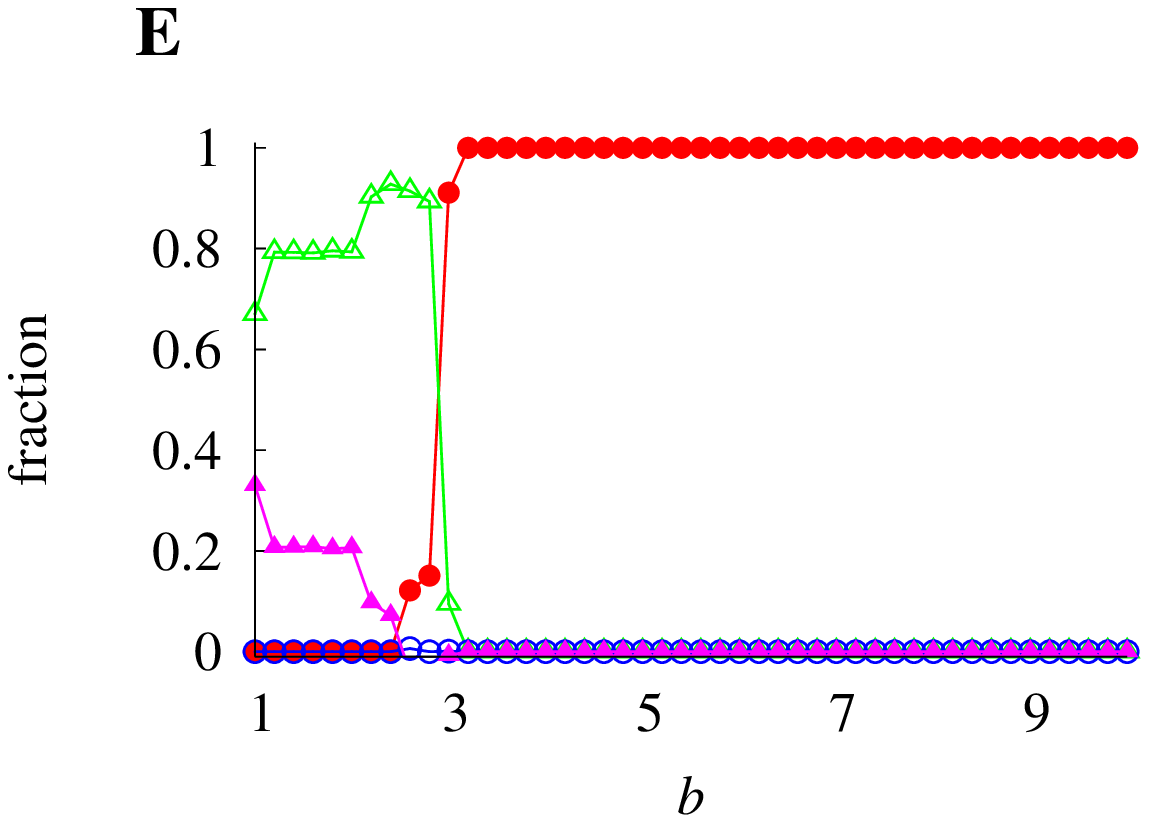}
\caption{Final fractions of four strategies
when players initially adopt either GC, CD, CC, or PO in
(A) scale-free network,
(B) regular random graph, (C) square lattice,
(D) extended cycle, and
(E) cycle. We set $\left<k\right>=8$, $p_i=0.8$, and $N_{\rm u}=200$.}
\label{fig:4 strategies}
\end{center}
\end{figure}

The final fraction of each strategy in the five networks is shown in
\FIG\ref{fig:4 strategies}.  In the scale-free network (\FIG\ref{fig:4
strategies}A), CD and PO do not survive for any value of
$b$. The fraction of GC increases with the value of $b$.  In the
regular random graph, the GC does not survive, and the network is
almost entirely inhabited by the least cooperative players,
\textit{i.e.}, CDs (\FIG\ref{fig:4 strategies}B). 
For GC to survive, the value of $b$ larger than 10, which
is the upper limit
of $b$ examined in \FIG\ref{fig:4 strategies}B, is required.
In the square
lattice (\FIG\ref{fig:4 strategies}C), the extended cycle
(\FIG\ref{fig:4 strategies}D), and the cycle (\FIG\ref{fig:4
strategies}E), GC takes over CD at a sufficiently large value
of $b$.  The lowest to highest
threshold value of $b$ above which the GC survives follows the order of
the scale-free network,
the cycle, the extended cycle, the square lattice, and the regular
random graph. 

The results are
robust against various changes of the model, such as
the value of $\left<k\right>$
(\FIGS\ref{fig:different k}C and D),
the value of $p_i$ (\FIGS\ref{fig:different p}C and D),
the value of $N_{\rm u}$
(\FIGS\ref{fig:different N_u}C and D), and
the update rule
(\FIGS\ref{fig:different update rules}C and D).
The results in this section including the robustness results
are consistent with those obtained for the
populations that comprise two strategies (\SECS\ref{sub:GC CD},
\ref{sub:GC CC}, and \ref{sub:GC PO}).

\section{Discussion}\label{sec:discussion}

We have shown that heterogeneous networks enhance cooperative behavior
in a game of upstream reciprocity. Based on the property of the
simple random walk on networks, chains of helping behavior traverse
hub players more often than players having small degrees. Then, hubs
tend to gain a larger payoff. The most cooperative strategy
(\textit{i.e.}, GC) is stable once it inhabits hubs, from where it spreads
to the entire network. From a quantitative point of view, the impact
of heterogeneous networks on enhancing altruism can be much more
than that of spatial reciprocity in most cases.
Our results are robust against variation in some parameters of the model
($\left<k\right>$, $p_i$, and $N_{\rm u}$) and variation in update rules.

The route to altruism in the game of upstream
reciprocity proposed
in this study is similar to that in the Prisoner's Dilemma on
heterogeneous networks \cite{Santos05prl,Duran05,Santos06pnas,Santos06jeb}. In this framework, each player is assumed to either cooperate with or
defect against all neighbors in a round.  Once a cooperator occupies
a hub and some surrounding nodes, the hub gains a
large payoff and is likely to disseminate its offspring
(\textit{i.e.}, cooperators) to the neighbors. This event further increases
the payoff of the hub, and the cooperation on the hub is
stabilized. In contrast, defection on a hub is not stable because the
hub does not gain a large payoff if the defector hub disseminates its
offspring to the neighbors. Cooperators are propagated from
hubs to the entire network. In the game of upstream reciprocity
in networks, suppose that a GC hub disseminates its offspring to the
neighbors. This hub will gain a larger
payoff in the subsequent rounds because the neighbors will tend
to pass on the chains of helping behavior.
Then, the GC hub will receive
helping behavior more often than typical players
such that its payoff increases, and the GC
is stabilized on the hub. This positive feedback is weaker in the case
of the PO and absent in the case of the CD
and CC.

When player $X$ with a small degree copies the strategy of a
successful hub neighbor $Y$, $X$ may not gain a large payoff because $X$
is not a hub.  In the Prisoner's Dilemma on networks, many previous
studies assumed that
selection is based on the summed payoff; in this,
each player sums up the payoff
obtained by playing against all neighbors to determine the payoff
per round \cite{Santos05prl,Duran05,Santos06pnas,Santos06jeb}.  However,
it may be advantageous for $X$ not to copy the strategy of $Y$, because $X$
is not as connected as $Y$. It may be more profitable for $X$ to
copy the strategy of a neighbor that earns a larger payoff per
edge. This update rule corresponds to the selection 
based on the average payoff,
\textit{i.e.}, the summed payoff divided by the degree.
The average payoff scheme does not enhance cooperation in the
Prisoner's Dilemma on heterogeneous networks
\cite{Santos06jeb,Tomassini07}. This argument is also applicable to
the game of upstream reciprocity in
scale-free networks. The evolution of
helping behavior is likely to be hampered if the selection is based on
average payoff. This is a major limitation of the present study.
The update rule that we have adopted, as well as the rule based on
additive payoff used in the
Prisoner's Dilemma, may represent a situation in which players are
unaware of the degree of their neighbors.

In the game of upstream reciprocity,
hubs gain relatively large payoffs 
because a
simple random walker visits hubs relatively often.  This is true for an
eternally lasting random walk on arbitrary undirected networks
\cite{Noh04prl}. However, in our model, the random walk 
terminates in finite time. Then,
the random walker may visit specific non-hub nodes more frequently
than it visits hubs, as in the case of the random walk in networks
with an absorbing boundary
\cite{Noh04prl,Newman05soc}. For heterogeneous networks in which
populations are not well mixed,
perhaps with degree correlation between adjacent nodes 
or global structure of networks,
our results may be modified. The GC may spread from
specific non-hub players. In directed
networks, the frequency of visit of the random walker to nodes can
also deviate from the predicted value based on the degree
\cite{Donato04,Masuda09njp}. Roughly speaking, however,
the random walk tends to visit more connected players
under all discussed cases. Therefore, 
we expect that our results qualitatively hold true for 
general heterogeneous networks.

\section*{Acknowledgments}
NM is supported by Grants-in-Aid for Scientific Research from MEXT,
Japan (Nos. 20760258, 20540382, and 20115009).

\newpage

\clearpage

\clearpage

\clearpage

\clearpage

\clearpage

\clearpage

\clearpage

\clearpage

\clearpage

\clearpage

\end{document}